\documentclass[12pt,preprint]{aastex}

\slugcomment{}
\usepackage{natbib}

\shorttitle{Triggered Star Formation in the Orion Region}
\shortauthors{H. Lee et al.}

\begin{document}

\title{TRIGGERED STAR FORMATION \\
    IN THE ORION BRIGHT-RIMMED CLOUDS}

\author{Hsu-Tai Lee\altaffilmark{1} \email{eridan@astro.ncu.edu.tw}}

\author{W. P. Chen\altaffilmark{1,2} \email{wchen@astro.ncu.edu.tw}}
  
\author{Zhi-Wei Zhang\altaffilmark{1}}

\and

\author{Jing-Yao Hu\altaffilmark{3}}

\altaffiltext{1}{Institute of Astronomy, National Central University, Taiwan 320, R.O.C.}

\altaffiltext{2}{Department of Physics, National Central University, Taiwan, 320, R.O.C.}

\altaffiltext{3}{National Astronomical Observatories, Chinese Academy of Sciences,
Beijing 100012, People's Republic of China.}

\begin{abstract}

We have developed an empirical and effective set of criteria, based on the 2MASS colors,
to select candidate classical T Tauri stars (CTTS).  This provides a useful tool to study 
the young stellar population in star-forming regions.  Here we present our analysis of the 
bright-rimmed clouds (BRCs) B\,35, B\,30, IC\,2118, LDN\,1616, LDN\,1634, and Orion\,East 
to show how massive stars interact with molecular clouds to trigger star formation.  Our 
results support the radiation-driven implosion model in which the ionization fronts from 
OB stars compress a nearby cloud until the local density exceeds the critical value, thereby 
inducing the cloud to collapse to form stars.  We find that only BRCs associated with 
strong IRAS 100~$\micron$ emission (tracer of high density) and H$\alpha$ emission (tracer 
of ionization fronts) show signs of ongoing star formation.  Relevant timescales, including 
the ages of O stars, expanding H\,{\small II} regions, and the ages of CTTS, are 
consistent with sequential star formation.  We also find that CTTS are only seen between 
the OB stars and the BRCs, with those closer to the BRCs being progressively younger.  There 
are no CTTS leading the ionization fronts, i.e., within the molecular clouds.  All 
these provide strong evidence of triggered star formation and show the major roles massive 
stars play in sustaining the star-forming activities in the region.

\end{abstract}

\keywords{stars: formation --- stars: pre-main sequence --- ISM: clouds --- ISM: molecules}

\section{INTRODUCTION}

Young stars tend to form in clusters or groups.  A giant molecular cloud may 
collapse and fragment to form stars of different masses.  The formation of 
massive stars has an immense impact on the environments.  On one hand 
stellar winds and shock waves from a supernova explosion may squeeze molecular 
clouds and induce subsequent birth of stars which otherwise may not have 
occurred.  On the other hand the agitation may be so violent as to 
disperse the material, hindering further star-forming activity.  Triggered 
star formation \citep[see][for a comprehensive review]{elm98} --- as 
opposed to spontaneous cloud collapse --- would have profound consequences 
in stellar population, chemical homogeneity, star formation efficiency, 
and cloud energetics.

While the OB and young low-mass member stars are often distributed 
cospatially, in some cases low-mass pre-main-sequence (PMS) stars are 
seen to be located between the molecular cloud and the cluster.  Such a 
configuration can result from either triggered or spontaneous star 
formation.  In the former case, high-mass stars form
in dense molecular cores, and then 
their UV photons create an expanding Str\"{o}mgren sphere.  The expanding 
ionization fronts (I-fronts) then compress nearby molecular clouds and 
trigger the formation of next-generation PMS stars at the edge of 
clouds.  Alternatively, star formation may proceed in a spontaneous way; 
that is, all stars form coevally but high-mass stars photoionize the 
surrounding clouds to expose the low-mass stars. 

The Orion complex is an active star-forming region with a wealth of 
star-forming signatures, such as high-mass star clusters, OB associations, 
low-mass young stars, giant molecular clouds, reflection nebulae and 
H\,{\small II} regions.  Because of its proximity, at $\sim$450~pc, it is a 
suitable laboratory for the study of star formation, especially for 
the interplay among high-mass stars, low-mass stars, and interstellar 
medium.

Bright-rimmed clouds (BRCs) are usually found at the boundary of 
H\,{\small II} regions and the edge of molecular clouds.  The morphology 
of a BRC, e.g., tightly curved or cometary rims with tails, is governed 
by the I-fronts from nearby OB stars, which compress the head of the BRC, 
hence making the density of the apex higher than the other parts of 
the cloud.  The imprints of such causality offer the possibility to 
delineate the intricate processes of sequential star formation.  
\citet{sug91,sug94} catalogue 89 BRCs associated with IRAS point sources 
in both northern and southern hemispheres, some of which are also 
associated with Herbig-Haro objects and molecular outflows.  These BRCs are 
potential sites where triggered star formation might have taken place.  
\citet{kar03}
present a multiwavelength study of triggered star formation in the 
W5 H\,{\small II} region.  They find some young stellar objects around W5 
and suggest that the expanding H\,{\small II} region has induced star 
formation in the surrounding molecular clouds.

This paper provides observational diagnosis to identify PMS stars from 
their characteristic near-infrared colors.  In particular, PMS stars 
associated with BRCs in the Orion star-forming region are used as a 
tool to chronologize the star formation history to show how massive 
stars trigger the star formation processes in the region.  

We describe in \S 2 our methodology in selection of the PMS sample 
in parts of the Orion and Monoceros star-forming regions.  In \S 3 
we discuss how our data lend supportive evidence of triggered star 
formation in the Orion BRCs.  The conclusions are summarized in \S 4.  

\section{SELECTION OF YOUNG STELLAR SAMPLE}

Weak-line T Tauri stars (WTTS) are known to emit copious x-ray emission, and a 
majority of them have been identified by optical spectroscopic observations 
of strong x-ray sources \citep[see][for a review]{fei99}.  In comparison, 
classical T Tauri stars (CTTS) are at earlier evolutionary stages, before 
much of the circumstellar material has been dispersed.  CTTS show 
characteristic strong IR excess emission, particularly at mid-infrared 
wavelengths, and H$\alpha$ emission (equivalent 
width $>$ 5~\AA).  The IR excess is believed to originate from heated 
circumstellar dust, whereas the H$\alpha$ emission come from an active 
chromosphere and the boundary layer between the young star and the fast 
rotating circumstellar disk.  
As young stars evolve and gradually clear away the surrounding 
material, they show different amounts of IR excesses, hence their 
near-infrared colors \citep{lad92}.  Our PMS candidates have been selected 
on the basis of their near-infrared colors. 
 
\subsection{2MASS AND OTHER DATA}

\begin{figure}
\includegraphics[angle=90,scale=0.8]{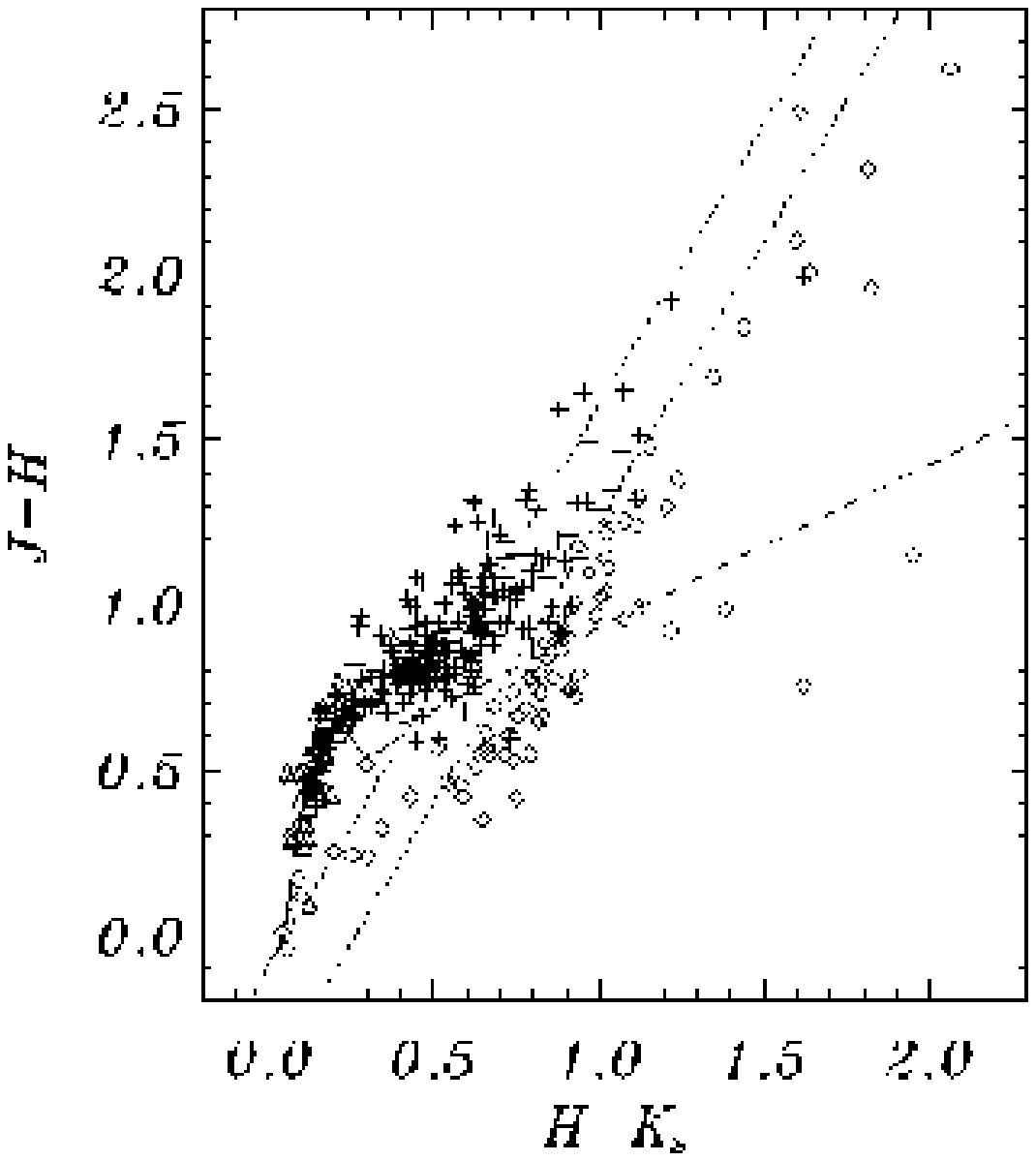}
\caption{The 2MASS $JHK$ color-color diagram for known WTTS (\citep{wic96}, 
triangles), CTTS (\citep{her88}, pluses) and Herbig Ae/Be stars 
(\citep{the94}, diamonds).  The solid line is the main
sequence locus.  The two dotted parallel lines, with the slope derived from interstellar
reddening law \citep{rie85}, separate CTTS from Herbig Ae/Be and from reddened main sequence stars.  
The dashed line is the dereddened CTTS locus \citep{mey97}.  
}
\label{fig:colors}
\end{figure}

This project started in early 2003, when we used the Two-Micron All-Sky Survey 
(2MASS) Second Incremental Data Release to identify CTTS candidates in the 
Orion and Monoceros regions.  Since the 2MASS 
released their All-Sky Data in March 2003, we have updated the photometry 
of our sample accordingly.  The 2MASS database provides unbiased photometry 
in the near-infrared $J$ (1.25~\micron), $H$ (1.65~\micron), and 
$K_{S}$ (2.17~\micron) bands to a limiting 15.8, 15.1, and 14.3~mag, 
respectively, with a signal-to-noise ratio (SNR) greater than 10.  
The comprehensive database offers a good opportunity to identify 
in a systematic way young star candidates associated with molecular clouds.  
We use known WTTS \citep{wic96}, CTTS \citep{her88} and Herbig Ae/Be stars 
\citep{the94} as our template to select PMS candidates.  
These lists of WTTS, CTTS and Herbig Ae/Be stars, taken from 
the literature and not confined to any special region in the sky,   
serve to the compilation of their descriptive 2MASS colors.  
Figure~\ref{fig:colors} plots the $(J-H)$ versus $(H-K_{S})$ color-color 
diagram, extracted from the 2MASS catalog for these PMS stars.  Only 
sources with high SNR photometry are included.  The fact that   
the CTTS, WTTS, and Herbig Ae/Be stellar populations appear to occupy 
different regions in Figure~\ref{fig:colors} enables us to pick out  
candidate PMS stars.  The 2MASS database, with usually 
simultaneous measurements in all 3 bands, is particularly useful for a 
diagnostic of young stellar population based on colors, because a majority 
of PMS stars are variables. 

Also shown in Fig.~\ref{fig:colors} 
is the locus for main sequence stars, for which the 
WTTS are seen to follow.  The CTTS are found to lie approximately between the 
two parallel dotted lines, defined empirically by $(j\_m-h\_m)-1.7(h\_m-k\_m)+
0.0976=0$ and $(j\_m-h\_m)-1.7(h\_m-k\_m)+0.450=0$,
where $j\_m$, $h\_m$, and $k\_m$ are 2MASS magnitudes, and the slope is 
specified by the interstellar reddening law \citep{rie85}.  
The dashed line in Fig.~\ref{fig:colors}  
represents the dereddened CTTS locus \citep{mey97}, modified to the 2MASS photometry 
\citep{car01}, with $(j\_m-h\_m)-0.493(h\_m-k\_m)-0.439=0$.  The dotted 
parallel lines separate CTTS from Herbig Ae/Be and from reddened main sequence stars. 
The fact that different kinds of PMS stars occupy separated regions in the $(J-H)$ 
versus $(H-K_{S})$ color-color diagram enables us to identify candidate young stars in 
star-forming regions.  We select 2MASS point sources with colors between the
2 parallel lines and above the dashed line as CTTS candidates.  
Inevitably our selection would miss some CTTS to the left of the parallel 
lines, where slightly reddened CTTS and reddened main sequence stars cannot 
be distinguished by their 2MASS colors.  This poses no difficulty for our study because 
our goal is not to collect a complete PMS sample, but to use young stars  
as probes of the star formation history in the region.  Data at longer wavelengths, 
e.g., at L band and beyond, would serve to better discriminate between moderately 
reddened CTTS and reddened main sequence stars \citep{lad00}. 

To improve photometry accuracy and to remove extended sources from the 2MASS Point 
Source Catalog, we have set further criteria in selection of our CTTS candidates.  
These may be useful to other users of the 2MASS data, so we summarize them in 
Table~1.   The Photometric quality flag (ph\_qual = AAA) means SNR $\geq$ 10, and with 
a corrected photometric uncertainty less than 0.10857~mag, an indication of the best 
quality detection in terms of photometry and astrometry.  The Blend flag (bl\_flg) 
= 111 implies no source blending; the Contamination and Confusion flag 
(cc\_flg) = 000 means the source is unaffected by known artifacts, or artifact
is not detected in the band.  Only sources without a neighboring star within 
5$\arcsec$ (prox $>$ 5.0$\arcsec$) are considered, so as to avoid possible 
photometric confusion.  We select only sources which have optical counterparts 
(a $\neq$ 0) and identified CTTS candidates by the near-IR colors prescribed in 
the last paragraph.  

Some slightly extended sources may be included in the 2MASS Point Source Catalog, 
notably nuclei of galaxies.  In order to single out these non-stellar sources, 
additional criteria have been imposed.  An extended source can be discriminated 
against a point source by the difference between its measured fluxes determined 
with a point-spread-function (PSF) fitting and with aperture photometry.  
Conceivably, an extended source measured by PSF photometry would lose some flux 
at the outer image halo.  The 2MASS entries, $[jhk\_m]$, are default magnitudes mostly 
derived from PSF fitting, whereas the entries, $[jhk\_m\_stdap]$, are standard 
aperture magnitudes derived from 4$\arcsec$ radius photometric aperture.  
Therefore the parameter $(j\_m-j\_m\_stdap)+(h\_m-h\_m\_stdap)+(k\_m-k\_m\_stdap)$ 
should be close to 0 for point sources and significantly non-zero 
for extended objects.  We set $(j\_m-j\_m\_stdap) + (h\_m-h\_m\_stdap) + 
(k\_m-k\_m\_stdap) \le 0.3$ as our selection criterion.  Finally, we choose 
sources with SNR greater than 30 ($[jhk]\_snr > 30$) for 
high-quality photometry.  These stringent criteria would bias against 
sources in crowded regions, but as shown below, prove to lead to a high success 
rate in selection of CTTS candidates.  

In addition to 2MASS, we have made use of other survey data, such
as the H$\alpha$ \citep{fin03, gau01}, $E(B-V)$ reddening \citep{sch98}, and IRAS
100~$\micron$ emission in the Orion star-forming region to trace, respectively, the
distribution of ionization fronts, cloud extinction and IR radiation with respect
to the spatial distribution of our sample of young stars.

\subsection{SPECTROSCOPIC OBSERVATIONS}

\begin{figure}
\epsscale{1.0}
\includegraphics[angle=90,scale=0.8]{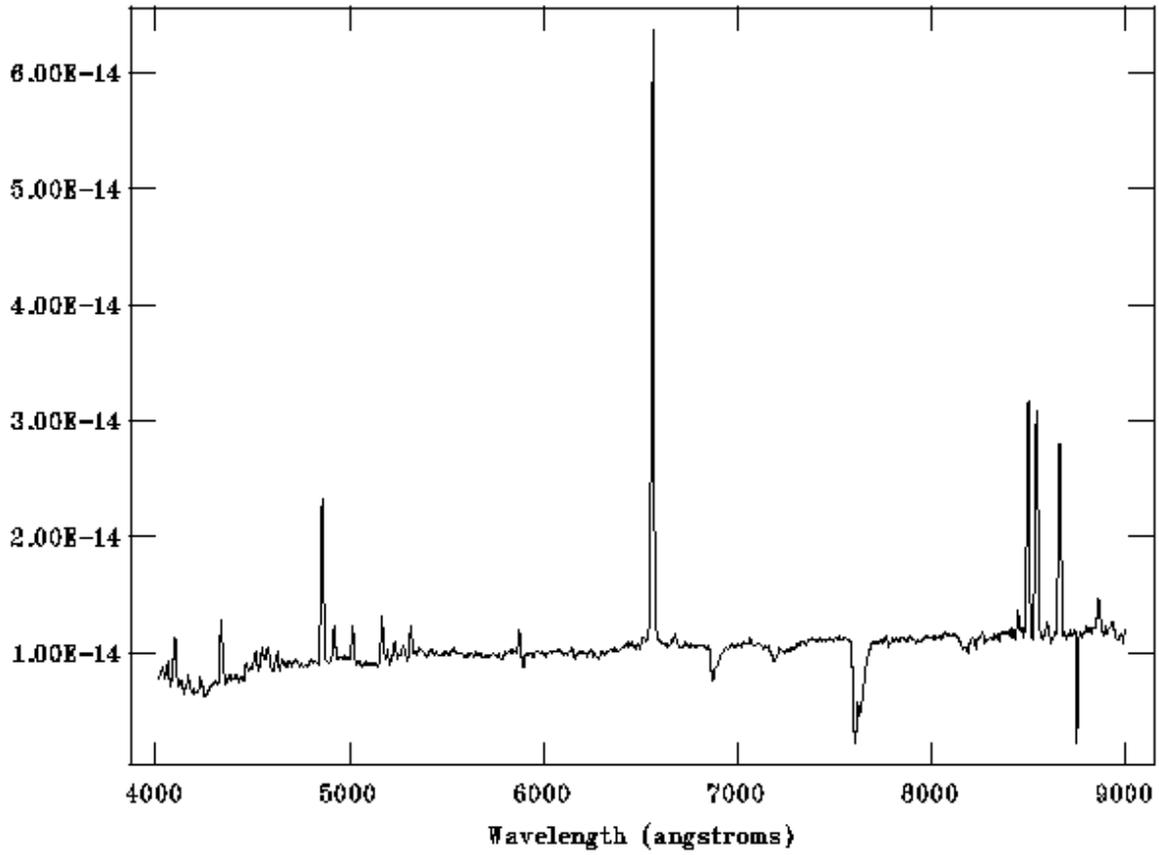}
\caption{Example CTTS spectrum for star No.\,16 showing extreme CTTS characteristics of 
a veiled continuum with strong H$\alpha$, Ca\,{\small II} triplets and forbidden  
[O\,{\small I}] and [S\,{\small II}] lines.}
\label{fig:spectrum}
\end{figure}

To justify the selection criteria outlined above, we have chosen 32 relatively 
bright CTTS candidates for spectroscopic observations to check their youth nature.  
Low-dispersion spectra, with dispersion of 200~\AA/mm, 4.8~\AA/pixel, were 
taken with the 2.16~m optical telescope of the Beijing Astronomical Observatory 
in 2003 January~22, 23, 30, and 31.  An OMR (Optomechanics Research Inc.) 
spectrograph was used with a Tektronix 1024$\times$1024 CCD detector covering 
4000-9000~\AA.  The data were processed with standard NOAO/IRAF packages.  
After the bias and flat-fielding corrections, the IRAF package KPNOSLIT 
was used to extract, and to calibrate the wavelength and flux of, each spectrum.  
Table~2 lists the results of our spectroscopic observations.  
The first column gives the sequence number.  Columns 2--5 list the 
2MASS identification and photometry.  Columns 6--9 are, respectively, the 
spectral type, the associated star-forming region, forbidden line(s) in the 
spectrum if any, and other names for the source.  We note that stars No. 3, 5,
6, 14, 15, 16, 20, 21, 22, and 24 in Table~2 show continuous or veiled spectra
with [O\,{\small I}] and/or [S\,{\small II}] emission lines
(Figure ~\ref{fig:spectrum}), characteristics of extreme T Tauri stars for
which the forbidden lines originate from jets or winds seen commonly in
Class~I sources \citep{ken98}.

These 32 candidates, distributed between RA$\sim$5--6~h covering part of 
the Orion and part of the Monoceros star-forming regions, have been 
chosen somewhat randomly.  Our spectroscopic observations show that 24 of 
the 32 candidates are found indeed to be pre-main sequence stars, with the other 
4 as M-type stars and 4 as carbon stars.  
All 24 confirmed PMS stars are associated with 
star-forming regions, but otherwise the M stars and carbon stars are scattered 
around the region.  This indicates that our criteria to select PMS 
stars from the 2MASS database are valid and very effective.  If the same 
set of criteria are applied to sources seen against nearby molecular clouds --- 
as is the case here in Orion and Monoceros bright-rimmed clouds ---  
for which both background and foreground field star contamination is small, 
the success rate should be conceivably even higher.  

\begin{figure}
\includegraphics[angle=90,scale=0.4]{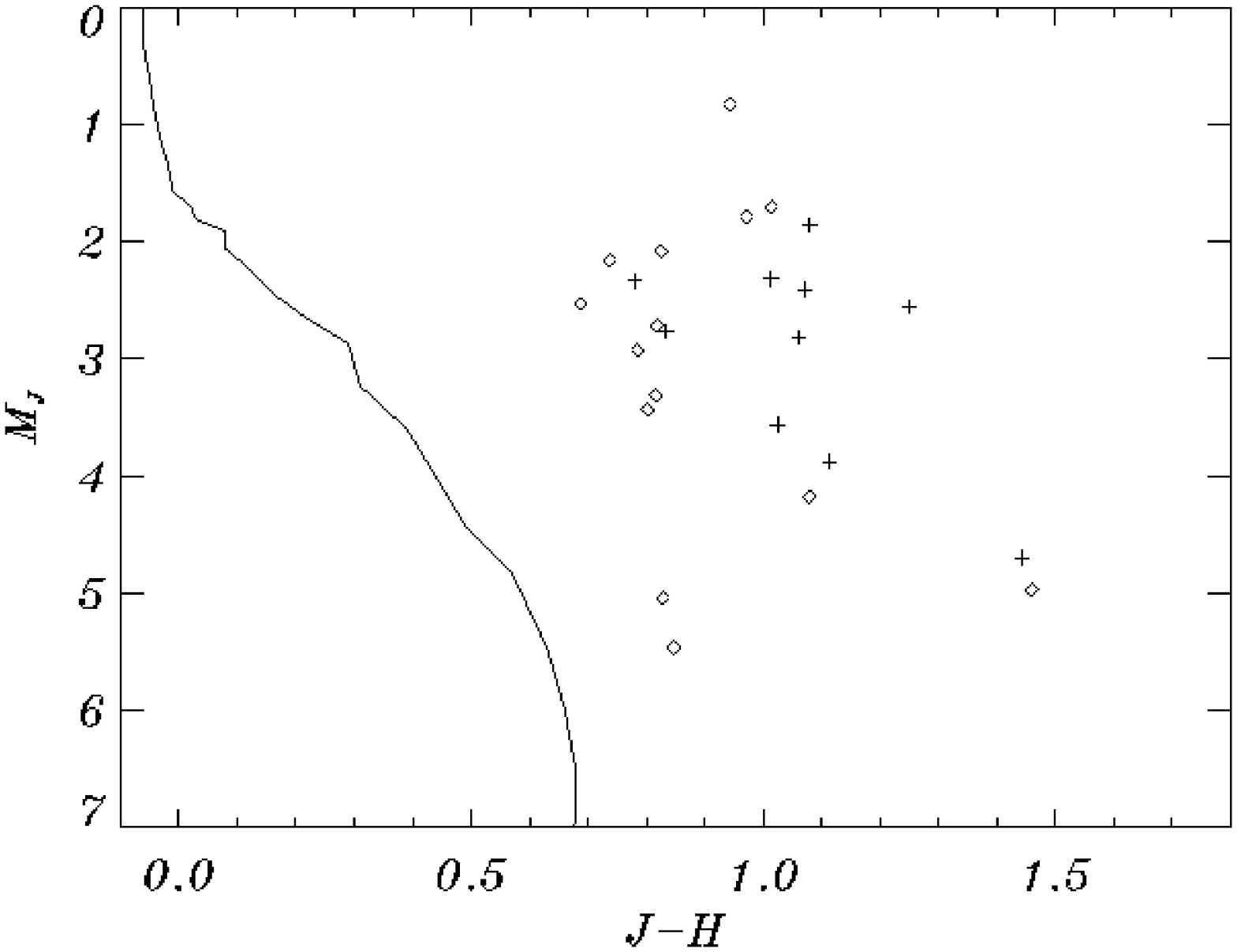}
\includegraphics[angle=90,scale=0.4]{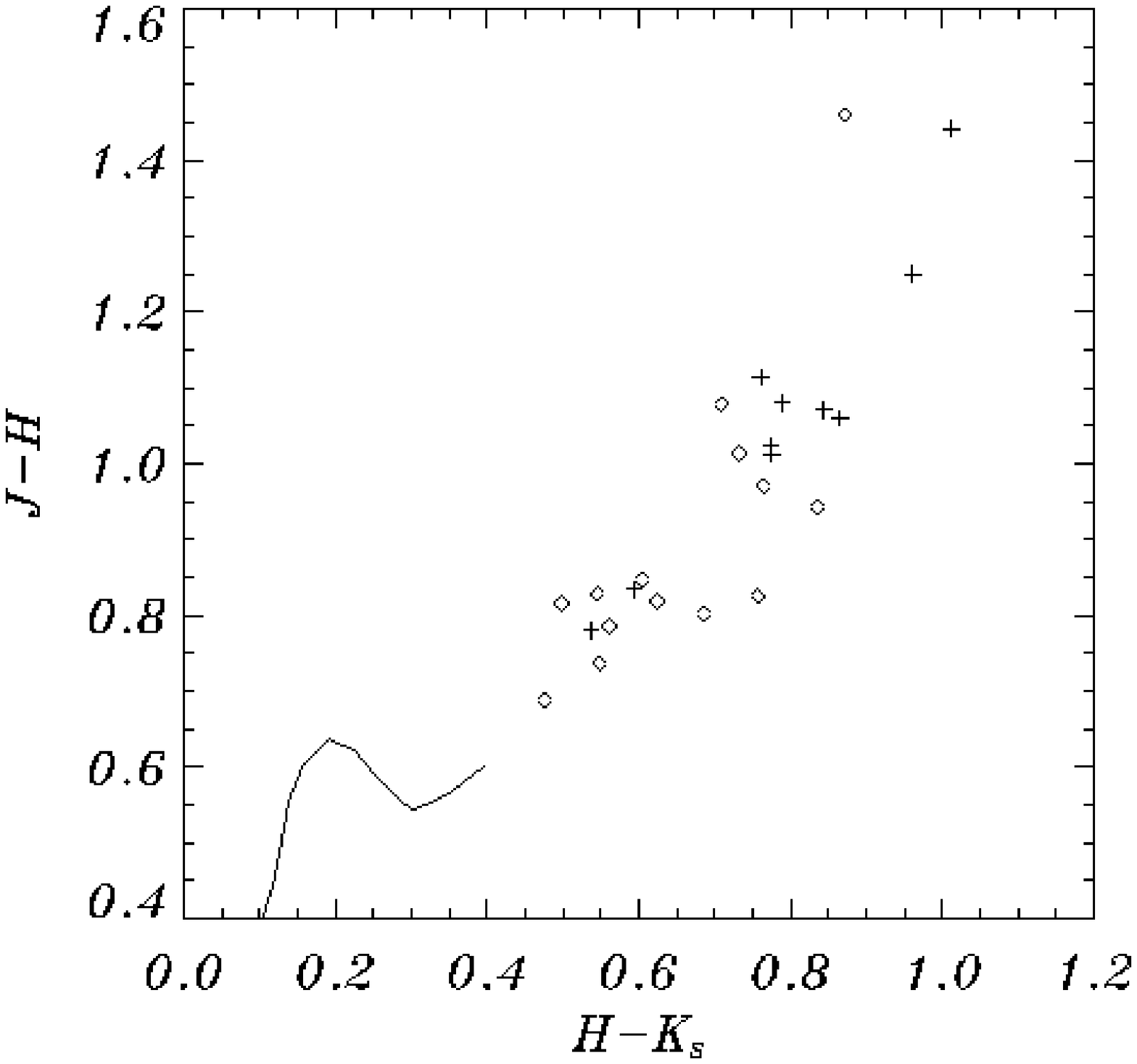}
\caption{Color-magnitude diagram and color-color diagram for the 24 verified 
CTTS in Table~2 with (pluses) and without (diamonds) optical forbidden lines.  
In the color-magnitude diagram 
the solid line represents the ZAMS.  In the color-color diagram the solid line is the main 
sequence locus.  In either case, the CTTS with forbidden lines are further way from 
the main sequence, suggestive of their redder colors and younger ages than those 
without forbidden lines.}
\label{fig:cmdforbidden}
\end{figure}

Figure~\ref{fig:cmdforbidden} shows the color-magnitude and color-color diagrams 
of the 24 spectroscopically confirmed CTTS in Table~2.   Stars with and without 
forbidden lines are denoted with different symbols.  In the color-magnitude 
diagram, the stellar 
absolute $J$ magnitude, as the ordinate, is derived by adopting a distance of 
450~pc to the Orion clouds and a distance of 830~pc to the Mon\,R2 and 
LDN\,1652 star-forming regions \citep{mad86}.  One sees that in either  
the color-magnitude diagram or the color-color diagram, the CTTS with 
forbidden lines are systematically further away from the zero-age main sequence 
(ZAMS), implying that the CTTS with forbidden lines are younger than those without.  

This result, that younger CTTS are located to the upper-right corner in 
the near-infrared color-color diagram, is consistent with that obtained 
by \citet{lad92}.   Class~I sources, with very red colors, $(J-H) > 1.8$~mag, 
are located even more toward the upper-right corner in the diagram.  In contrast, 
the WTTS are close to the main sequence locus, whereas the CTTS occupy a region 
between those of the Class~I sources and the WTTS.  This is understood as an 
evolutionary sequence; namely a young star evolves in such a color-color 
diagram from upper right to lower left, in a sequence from a Class~I 
protostar, a CTTS with forbidden lines, a CTTS without forbidden, to a WTTS.    

The spectroscopic observations demonstrate that our selection of PMS sample  
based on 2MASS colors is reliable, and that the color-magnitude diagram 
and color-color diagram can be used, particularly for those 
CTTS too faint to obtain their spectra, to diagnose their evolutionary 
status.  We are now ready to apply these tools to identify a sample of PMS 
stars in star-forming clouds, which in turn allows us to probe the star formation 
activity and history in the region.   


\section{EVIDENCE OF TRIGGERED STAR FORMATION}

\begin{figure}
\includegraphics[angle=0.0,scale=0.9]{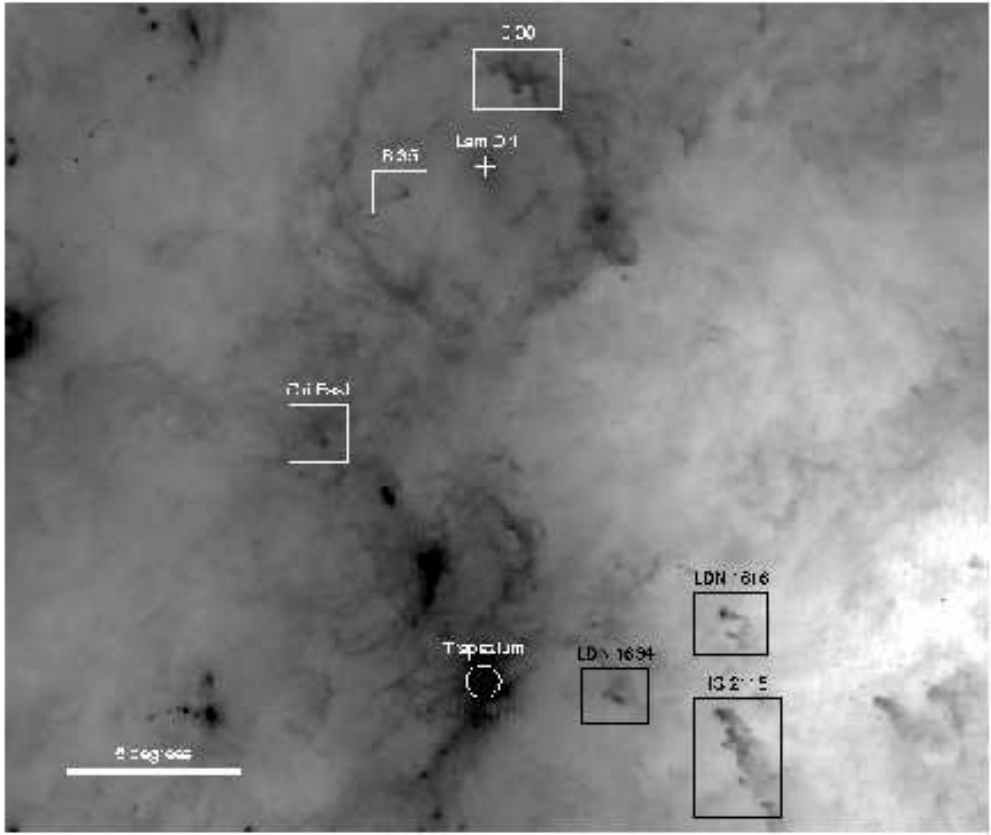}
\caption{IRAS 100~$\micron$ image of the Orion region with the 6 bright-rimmed 
clouds studied
in this work, B\,30, B\,35, Orion\,East, LDN\,1616, LDN\,1634, and IC\,2118.}
\label{fig:map}
\end{figure}

We present the analysis of the PMS population in relation with the 
bright-rimmed molecular clouds and massive stars to investigate how triggered star 
formation might have taken place.  This paper concentrates on the Orion 
BRCs, IC\,2118, LDN\,1616, LDN\,1634, and perhaps also Orion\,East, associated 
with the Trapezium, and B\,30 and B\,35 associated with the $\lambda$~Orionis 
(Figure ~\ref{fig:map}).  The sample used in this study consists of 
(1)~33 CTTS candidates (including 2 confirmed CTTS in IC\,2118 and 1 in 
LDN\,1634, see Table~2) around the Trapezium, in the region RA$\sim 5$~h 
to 5h\,30m, and DEC$\sim -2\deg$ to $-9\deg$, and (2)~18 CTTS candidates 
around $\lambda$ Orionis, in RA$\sim$5h25m to 5h50m, and DEC$\sim +8\deg$ to  
$+14\deg$. 

\begin{figure}
\includegraphics[angle=270,scale=0.5]{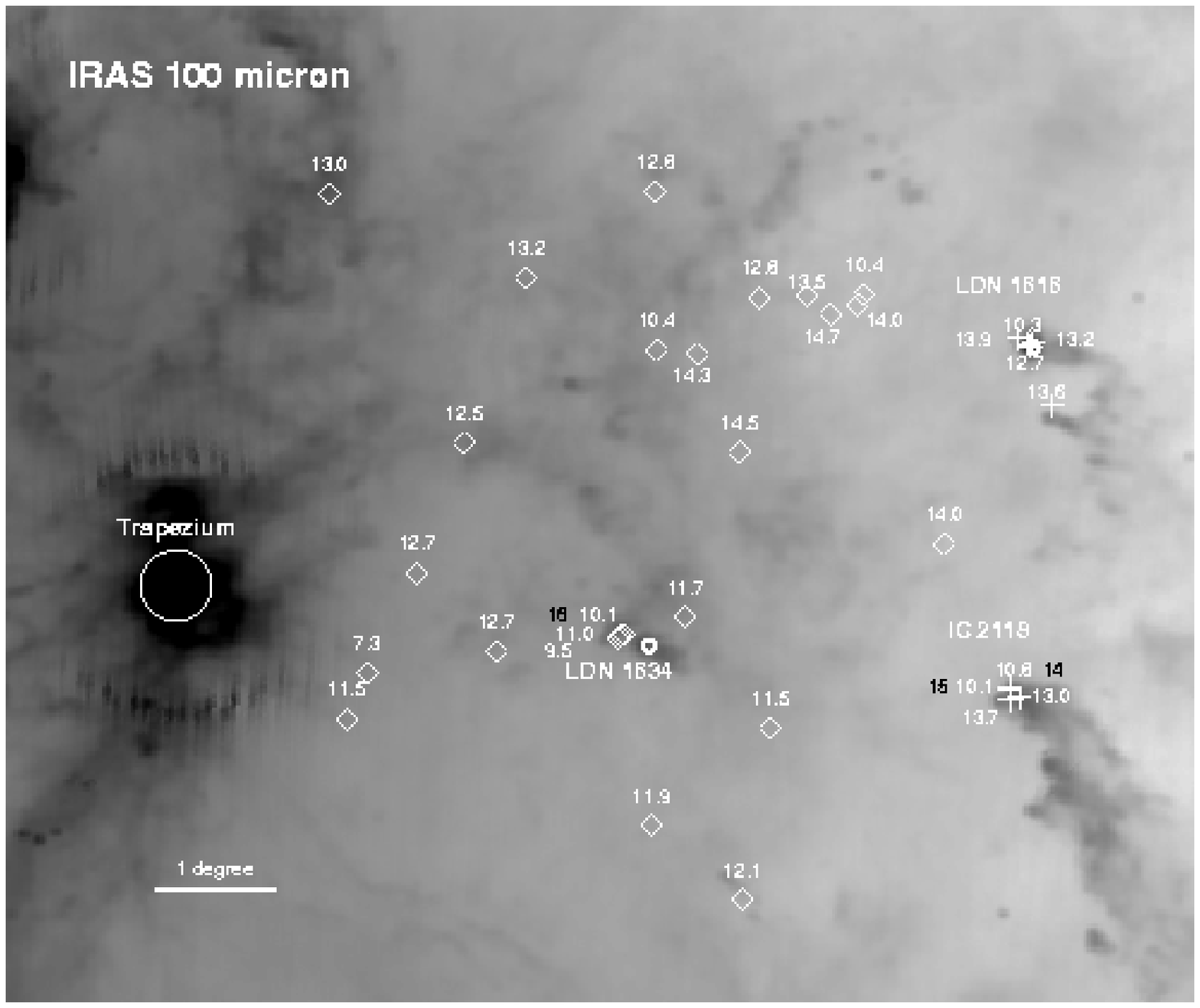}
\includegraphics[angle=270,scale=0.5]{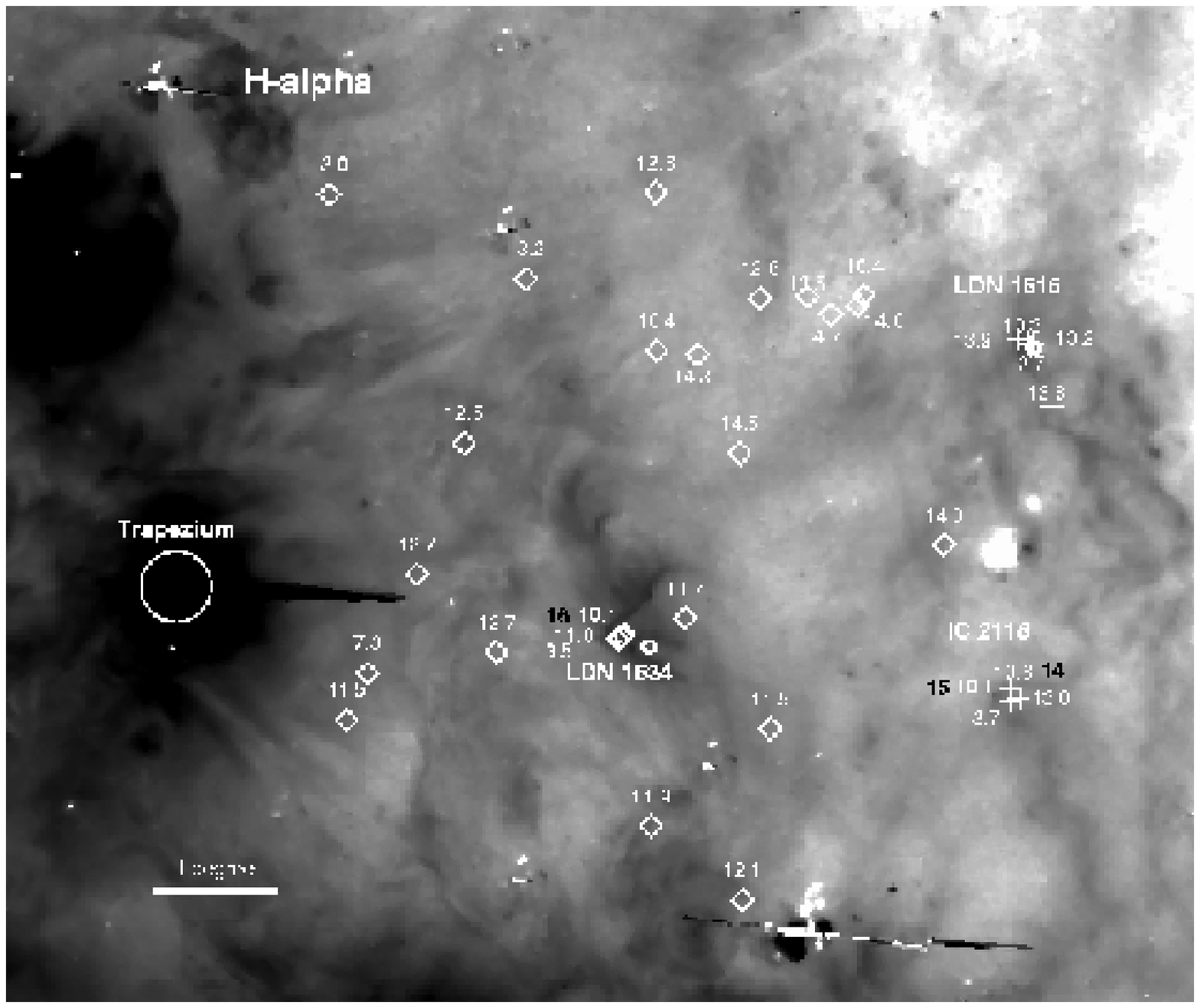}
\caption{IRAS 100~$\micron$ and H$\alpha$ images in IC\,2118, 
LDN\,1616, and LDN\,1634.  There are a total of 33 CTTS candidates in the region.
The pluses denote those CTTS physically closer to the BRCs, and 
the diamonds are those further away, with the J-band magnitude labeled for 
each star.  Stars No.~14, 15 and 16 in Table~2 are labeled in black.
The circles mark the positions of known protostars in the clouds (see \S 3.2).}
\label{fig:images5}
\end{figure}

IC\,2118, LDN\,1616, and LDN\,1634 are isolated molecular clouds to the 
west of the Orion\,A, as part of the Orion star-forming region.  
The 3 molecular clouds are all associated with strong IRAS 100~$\micron$ 
emission, and are apparently engraved by the UV radiation from the Trapezium, 
with the bright-rimmed boundaries, as outlined by H$\alpha$ filaments, presumably 
shaped by the I-fronts (Figure~\ref{fig:images5}).  We differentiate the CTTS
physically close to the BRCs from those further away, as marked in 
Figure~\ref{fig:images5}.   
The spectra of stars No.~14, 15 and 16, associated with IC\,2118 and 
LDN\,1634, all exhibit forbidden lines [S\,{\small II}] and/or [O\,{\small I}], 
suggestive of their youth.  The spatial distribution of these CTTS likely 
traces a pre-existent molecular cloud complex for which IC\,2118, LDN\,1616, 
and LDN\,1634 are merely survivors.  

\begin{figure}
\includegraphics[angle=270,scale=0.5]{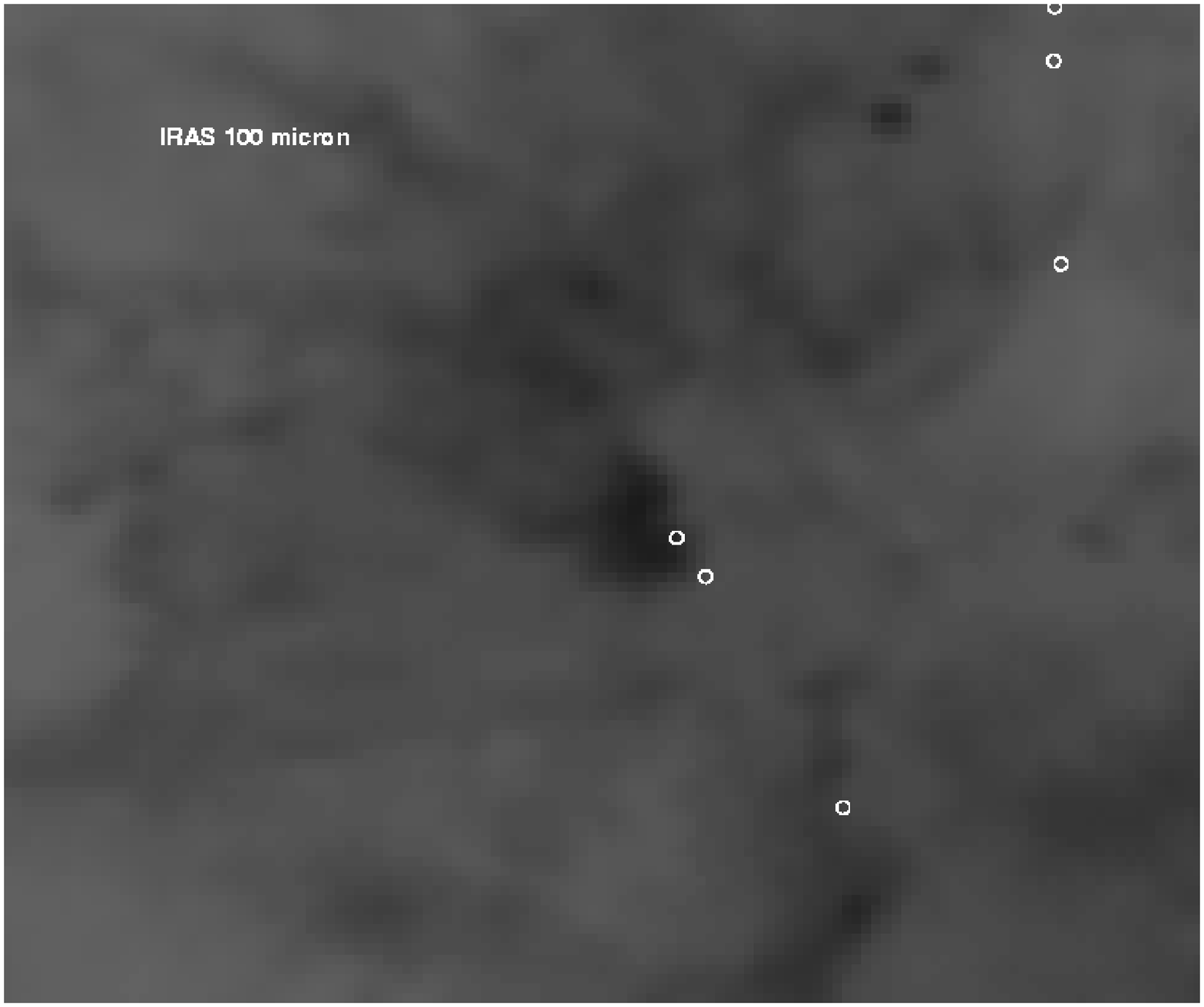}
\includegraphics[angle=270,scale=0.5]{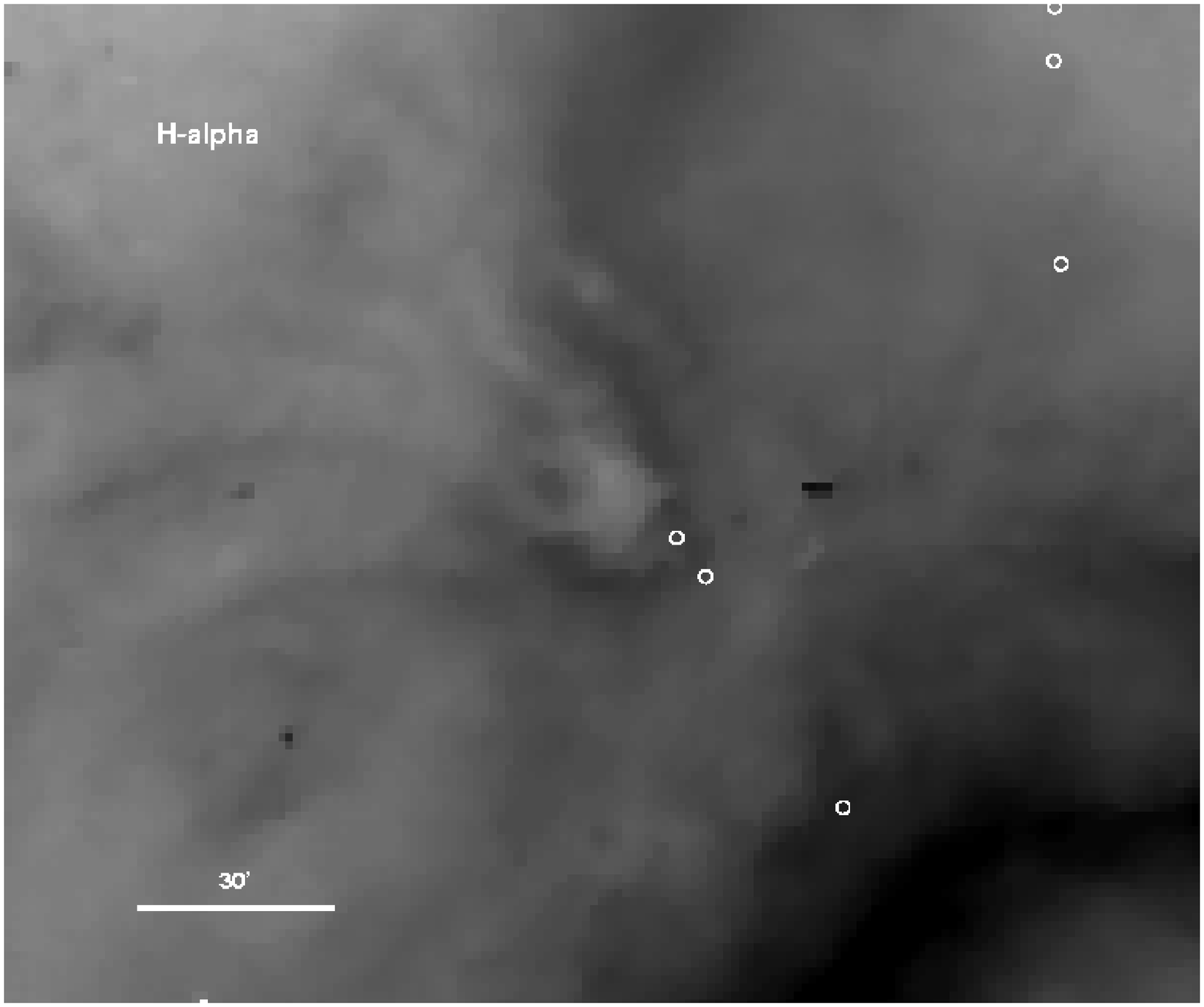}
\caption{IRAS 100~$\micron$ and H$\alpha$ images of Orion\,East, 
with candidate CTTS marked.}
\label{fig:images6}
\end{figure}

LDN\,1621 and LDN\,1622, also called Orion\,East \citep{her72}, are at 
about the same distance as Orion\,B \citep{mad86}.  Interaction between the 
Orion\,East clouds and the I-front can be seen in Figure~\ref{fig:images6}.

\begin{figure}
\includegraphics[angle=270,scale=0.5]{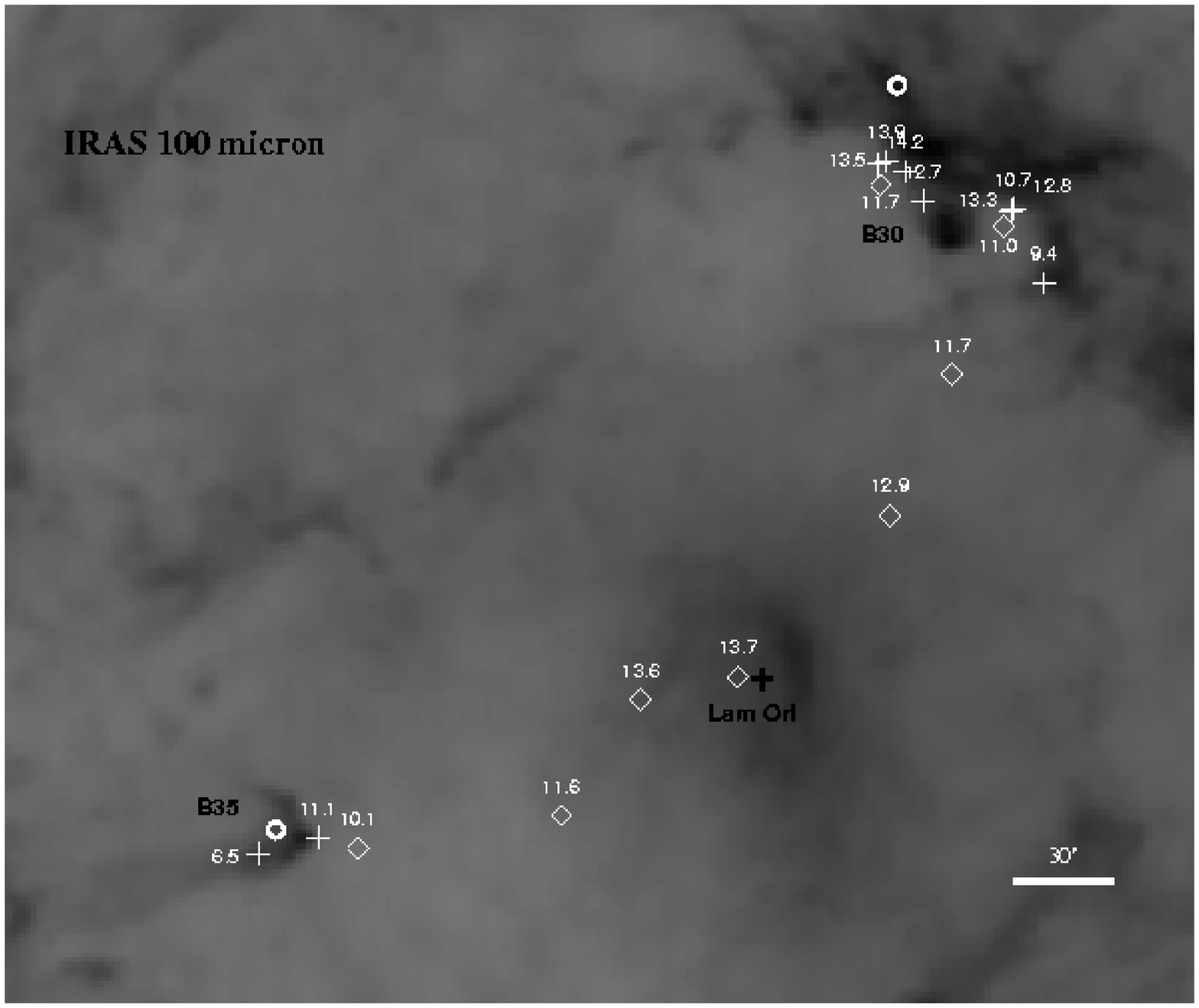}
\includegraphics[angle=270,scale=0.5]{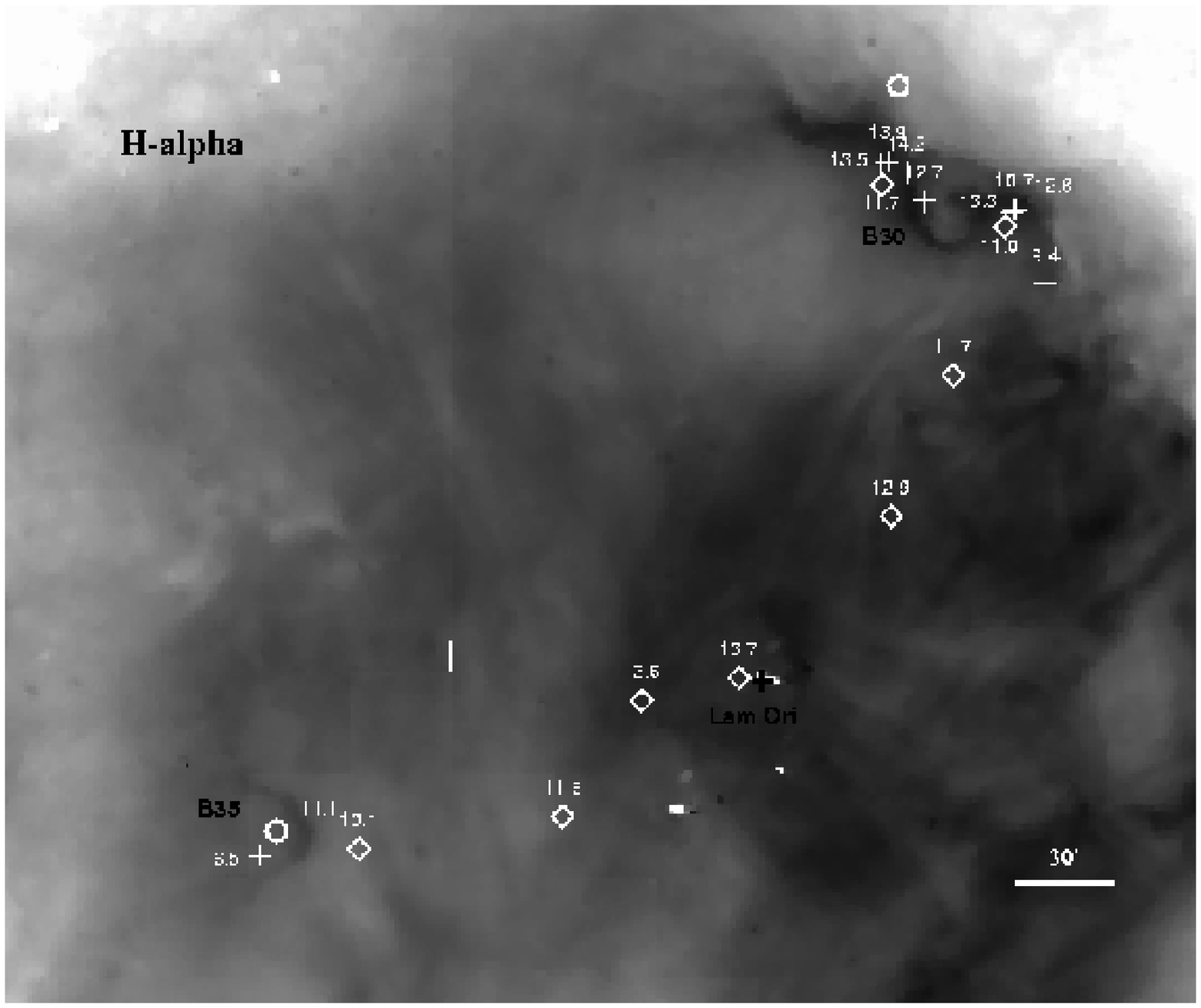}
\caption{Same as Fig.~\ref{fig:images5} but for B\,30 and 
B\,35.  There are a total of 18 CTTS candidates in the region.
}
\label{fig:images7}
\end{figure}

B\,30 and B\,35 are associated with S\,264, an extended H\,{\small II} region excited by 
the O8\,{\small III} star $\lambda$ Orionis and surrounded by, but slightly off centered 
of, a ring molecular cloud \citep{lan00}.  \citet{due82} have 
surveyed about 100 square degrees around $\lambda$ Orionis, and found 83 H$\alpha$ 
emission-line stars, mostly distributed along a bar-like structure extending 
on either side from $\lambda$\,Orionis to B\,30 and to B\,35.  
These authors suggest that the H$\alpha$ stars provide a fossil record 
of a pre-existent giant molecular cloud complex.  \citet{dol02} 
have identified PMS stars in the region around $\lambda$ Orionis by 
statistical removal of field stars in the optical color-magnitude diagram.  
The seeming lack of the youngest (1-2 Myr) PMS stars inside the molecular 
ring, despite the apparent abundance of such stars in B\,30 and B\,35, leads 
\citet{dol02} to postulate a possible supernova explosion near $\lambda$ Orionis 
that terminated recent star formation in the vicinity.  
The CTTS candidates we have identified spread along the bar-like structure, 
but clearly within the central cavity previously thought to be free of ongoing  
star-forming activities.  Prominent H$\alpha$ and IRAS 100~$\micron$ emission 
are seen in both B\,30 and B\,35 on the sides facing $\lambda$\,Orionis 
(Figure~\ref{fig:images7}), 
signifying the interaction between $\lambda$\,Orionis and these two BRCs.

\subsection{COMPARISON WITH THEORY}

The radiation-driven implosion (RDI) model, based on the ``rocket effect'', 
proposed by \citet{oor55} and further developed by \citet{ber89} and by 
\citet{ber90}, links BRCs to PMS stars.  The UV photons from massive stars ionize 
the outer layers of a nearby cloud, which expand to the surrounding medium with 
an I-front speed $\sim$ 10~km~s$^{-1}$.  The expanding I-front plows the cloud 
material which, when exceeding the local critical mass, collapses to form 
new stars.  
There are two phases, namely collapsing and cometary phases, for the 
formation and evolution of a cometary globule \citep{lef94,lef95,lef97}.  
In the initial collapsing phase, the UV photons ionize the surface layers 
of a cloud, causing the cloud to elongate in the direction of the high-mass 
stars.  This phase lasts some 
10$^{5}$~yr.  Subsequently, the cloud would remain in a quasi-stationary state, 
with small-scale condensations close to critical equilibria.  If the I-front 
out-pressures the molecular cloud, the density increases to beyond the 
critical value and the collapse of the condensations leads to star 
formation, leaving behind a chain of PMS stars reminiscent of the original, 
elongated cloud.  On the other hand, if the I-front does not out-pressure 
the molecular gas, the I-front stalls and there should be no star forming.  
This cometary phase may last a few million to a few 10$^{7}$~years.  
Eventually the entire cometary globule would be evaporated by the UV photons. 

According to the RDI model, the surface of a BRC facing the massive stars 
is compressed by the I-fronts. \citet{dev02} observed B\,35 and LDN\,1634 
in HCO$^{+}$ and other molecular line transitions to trace the swept-up gas 
ridge and overly dense regions.  They find that the surfaces of B\,35 and LDN\,1634 
facing the massive stars are indeed being compressed, as the RDI model predicts.

\begin{figure}
\includegraphics[angle=270,scale=0.5]{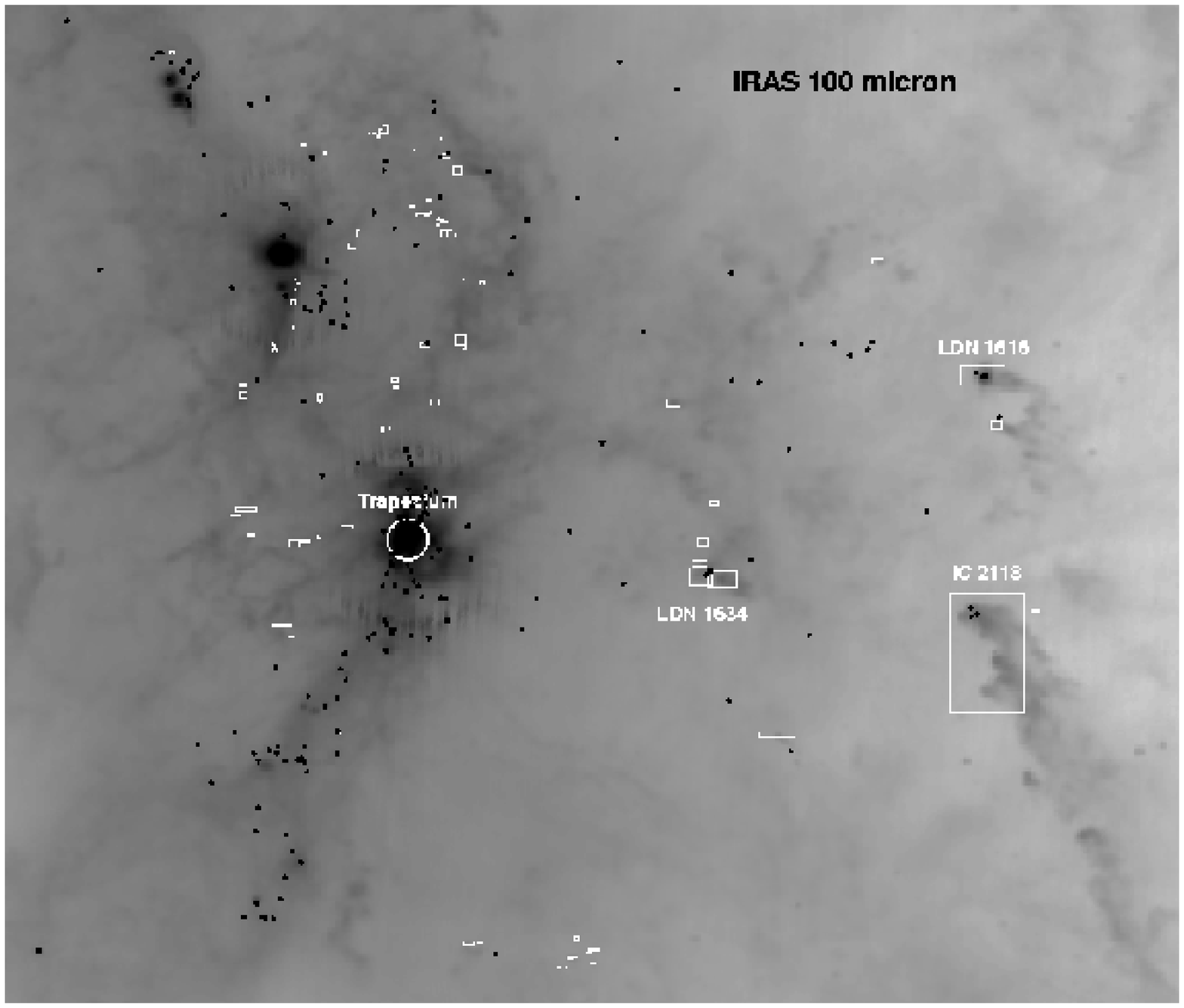}
\includegraphics[angle=270,scale=0.5]{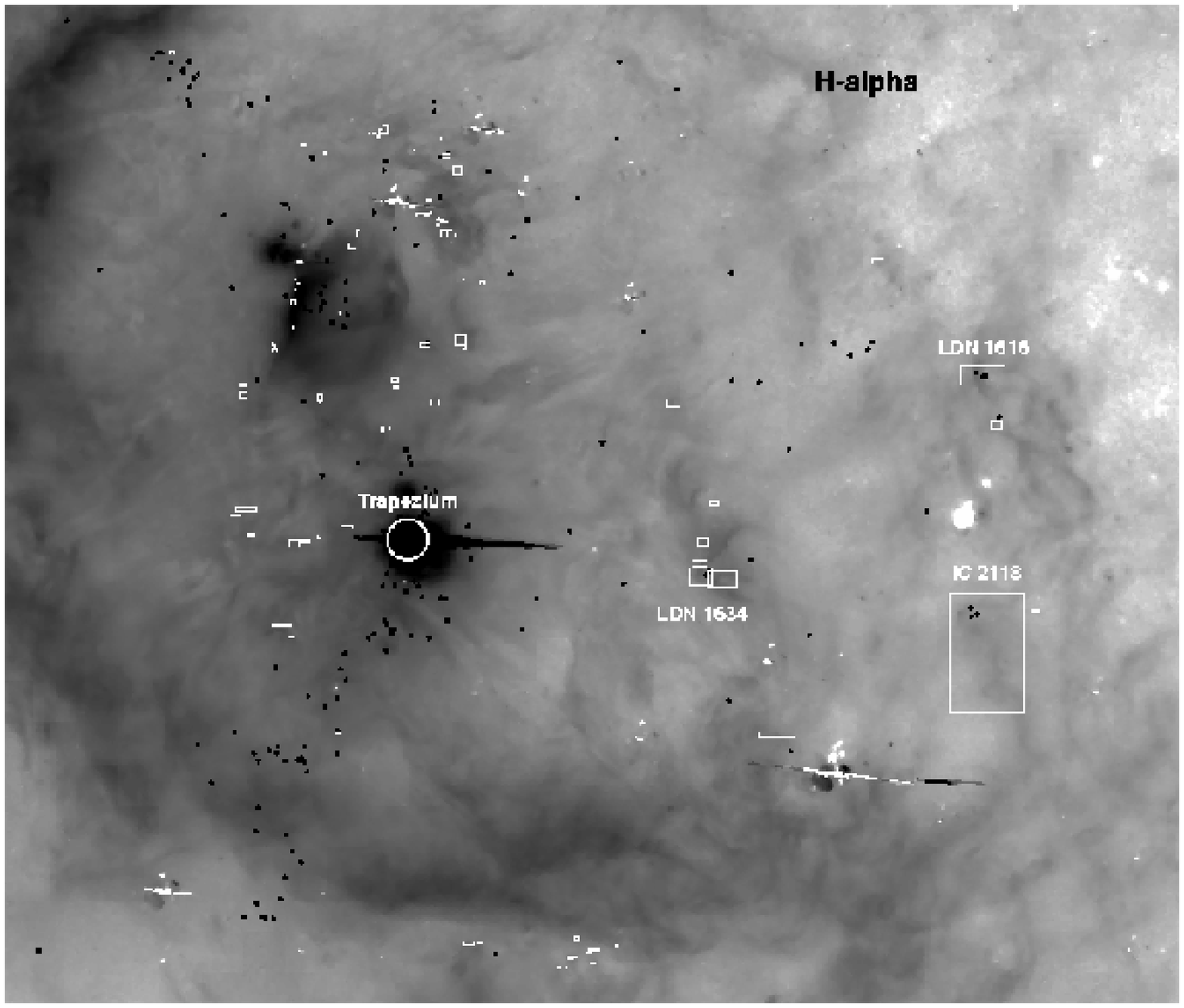}
\caption{IRAS 100~$\micron$ and H$\alpha$ images of remnant molecular clouds. 
The white boxes are the remnant molecular clouds \citep{ogu98}, and the 
black dots symbolize the candidate CTTS.  Only clouds associated with 
strong IR and H$\alpha$ emission are found to harbor CTTS.}
\label{fig:images8}
\end{figure}

\citet{ogu98} listed several dozens of bright-rimmed clouds, cometary globules 
and reflection clouds in the Orion\,OB1.  They propose these present an 
evolutionary sequence and collectively call them ``remnant molecular clouds''.  
We find that not all remnant molecular clouds are associated with CTTS.  
Instead, only those BRCs associated with prominent H$\alpha$ and IRAS 100~$\micron$ 
emission, e.g., IC\,2188, LDN\,1616, and LDN\,1634, have ongoing star formation 
(Figure~\ref{fig:images8}), 
consistent with the RDI model.  The H$\alpha$ filaments and the 
100~$\micron$ emission trace, respectively, the I-fronts and dense regions in 
a molecular cloud.  Obviously in IC\,2188, LDN\,1616, and LDN\,1634, the I-front 
has out-pressured the molecular clouds and thus triggered star formation.  The two 
BRCs B\,30 and B\,35 in the vicinity of $\lambda$~Orionis are additional two such 
examples of triggered star formation.  Remnant clouds without star formation 
are simply the outcome of I-fronts with insufficient pressure.

The RDI model predicts the cloud to collapse perpendicular to the direction 
of the ionizing massive stars.  This has been observed in LDN\,1616 and IC\,2118 
by \citet{yon99}, for which the $^{12}$CO, $^{13}$CO, and C$^{18}$O emission 
contours show elongation roughly toward the Trapezium.

\subsection{SEQUENTIAL STAR FORMATION}

Triggered star formation is a sequential process.  Massive stars form first, and 
their I-fronts expand to and compress nearby molecular clouds, prompting 
the next-generation stars to form.  The process may continue until the cloud is 
consumed and no more 
star-forming material is available.  As time passes, the PMS stars would line up 
between the massive stars and the molecular cloud, along which the stars farther 
away from the massive stars, and thus closer to the molecular cloud, should be 
progressively younger.  The youngest stellar objects, namely protostars 
(Class~0 and Class~I objects, \citealp{lad87}), would still be embedded within the 
cloud. 

\begin{figure}
\includegraphics[angle=90,scale=0.4]{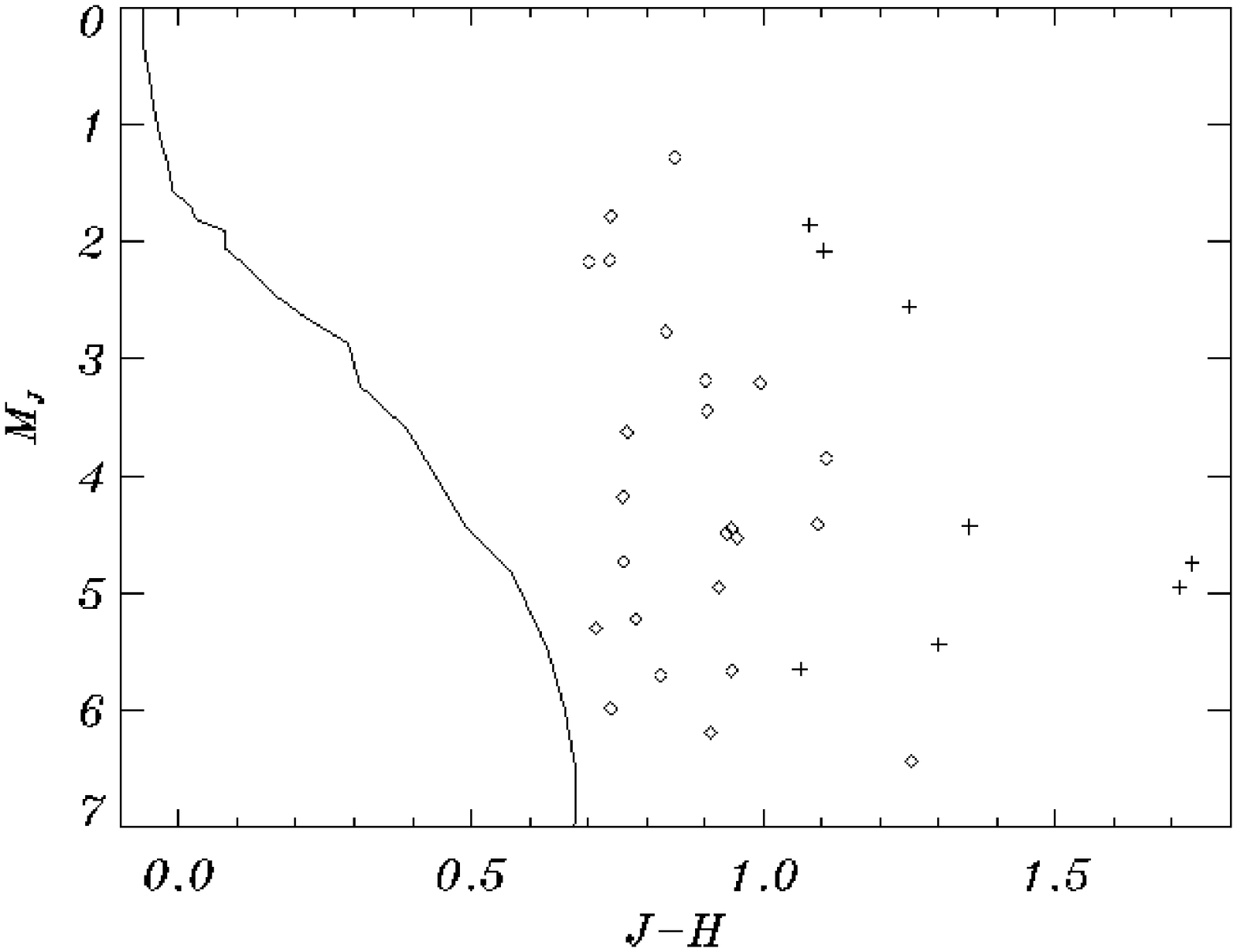}
\includegraphics[angle=90,scale=0.4]{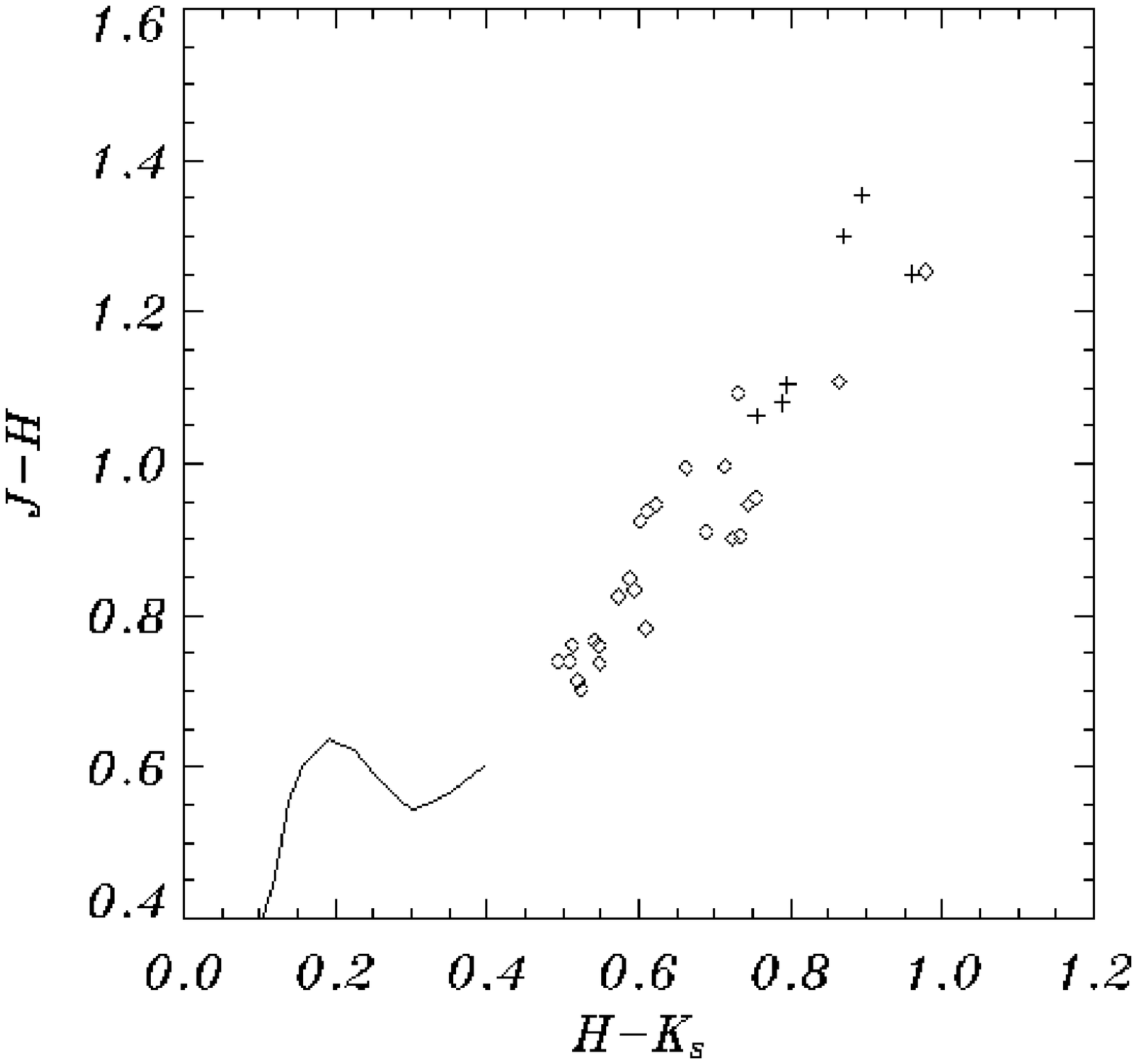}
\caption{Color-magnitude and color-color diagrams of IC\,2118, LDN\,1616, and LDN\,1634.  
In the color-magnitude diagram the solid line represents the ZAMS.  In the color-color 
diagram the solid line is the main sequence locus.
The CTTS physically closer to BRCs 
are marked with pluses, and 
those further away are marked with diamonds (cf. Fig.~\ref{fig:images5}). 
}
\label{fig:IC2118cc}
\end{figure}

\begin{figure}
\includegraphics[angle=90,scale=0.4]{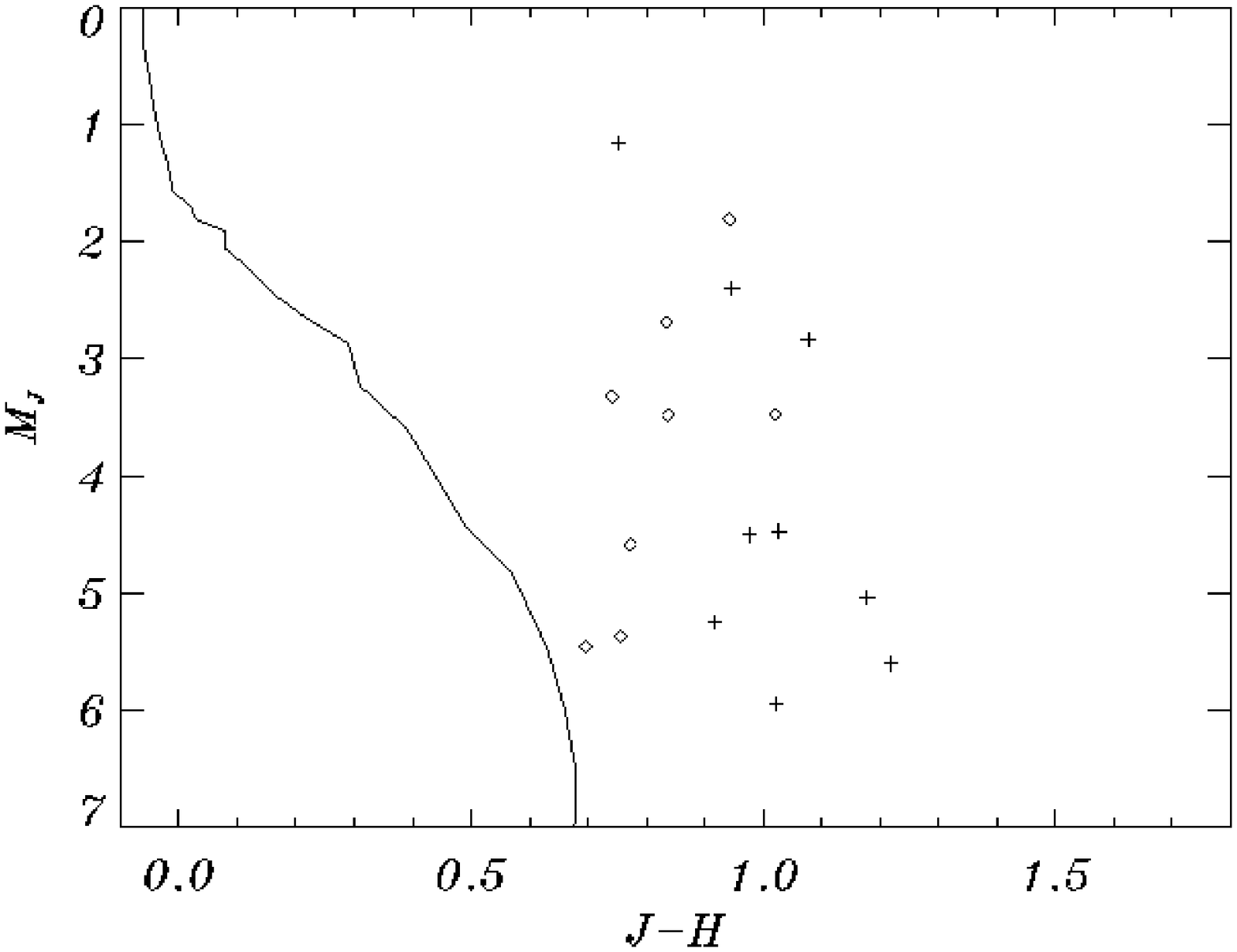}
\includegraphics[angle=90,scale=0.4]{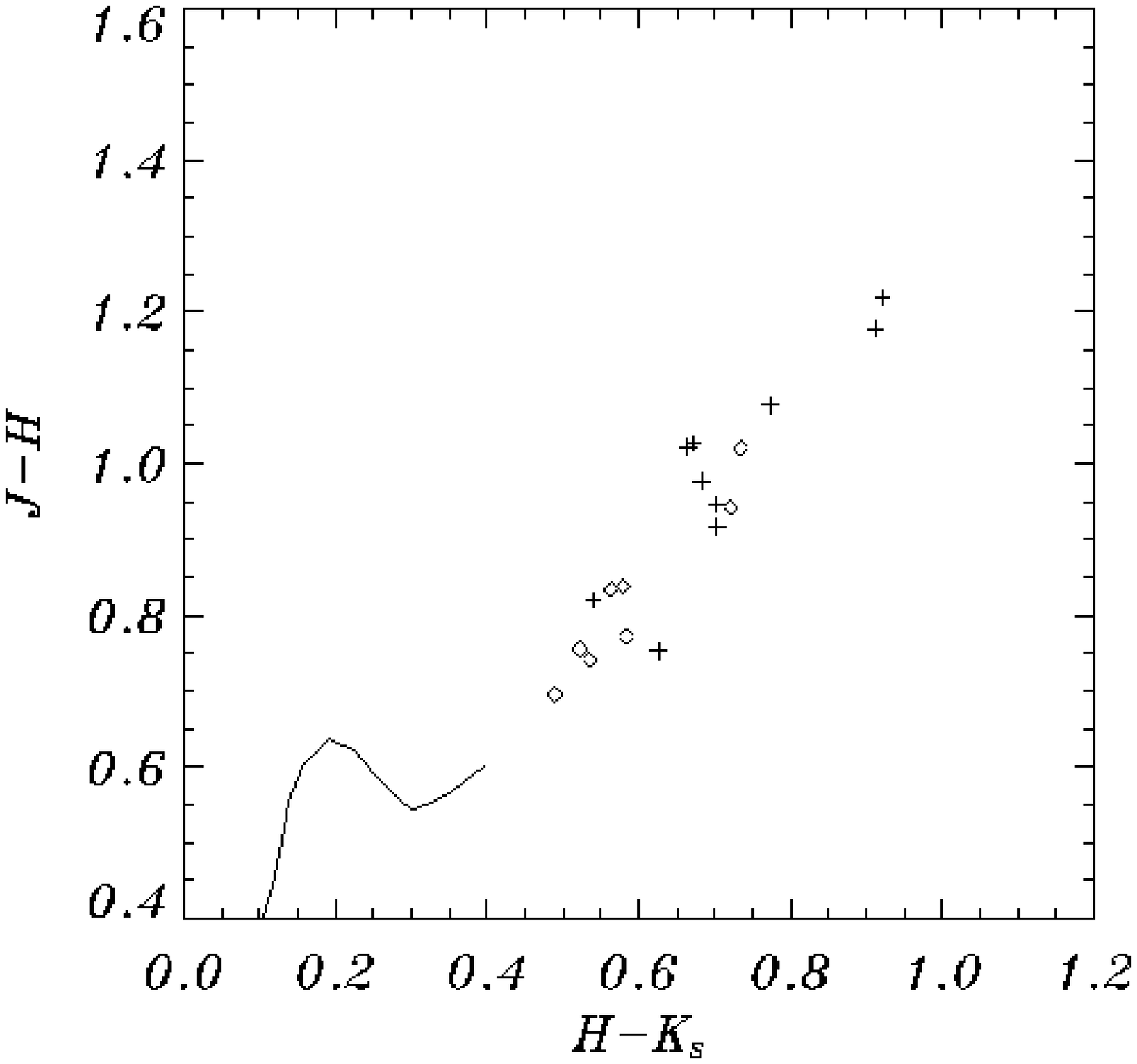}
\caption{Same as Fig.~\ref{fig:IC2118cc} but for the B\,30 and 
B\,35 region (cf. Fig.~\ref{fig:images7}.) }
\label{fig:Lamoricc}
\end{figure}

Our data indicate a systematic brightness difference among revealed CTTS, in 
the sense that brighter CTTS are more distant from $\lambda$~Orionis 
(cf. Fig.~\ref{fig:images7}).  
We consider this as the consequence of triggered star formation.  Initially, 
the B\,30 and B\,35 clouds might have been extended toward $\lambda$~Orionis, 
perhaps with a bar-like structure \citep{due82}.  The I-front from 
$\lambda$~Orionis propagates and rams through the clouds, leaving behind newborn stars.  
Figure~\ref{fig:Lamoricc} plots the color-magnitude diagram in the 
$\lambda$~Orionis region.  Indeed the CTTS closer to B\,30 and B\,35 are progressively 
younger, and away from zero-age main sequence \citep{sie00}.  
The brightest CTTS, close to B\,30 and 
B\,35, may be among the youngest which have just revealed themselves 
on the birthline and begun their descent down the Hayashi tracks.  
The same phenomenon, namely the CTTS closer to the BRCs are 
perceptibly younger, is seen in LDN\,1616, LDN\,1634, and IC\,2118 
(Fig.~\ref{fig:IC2118cc}).  
The star formation sequence is similar, except here with the Trapezium 
as the triggering source.

Figure~\ref{fig:Lamoricc} also shows the color-color diagram of the $\lambda$~Orionis 
region, in which the CTTS physically closer to BRCs are located to the upper-right corner.  
The extinction values in these BRCs, derived from \citet{sch98}, are in general too 
low to affect the apparent colors significantly.  This implies these CTTS are 
intrinsically young.  In LDN\,1616, LDN\,1634, and IC\,2118, we also find similar  
results as in the $\lambda$~Orionis region (cf. Fig.~\ref{fig:IC2118cc}).

Some very young objects are known to exist in these regions, for example with Class~0 
sources in B\,30 (RNO~43MM, \citealp{zin92}) and in LDN\,1616 (L\,1616~MMS1A, \citealp{sta02}), 
and with Class~I sources in B\,35 and in LDN\,1634 \citep{dev02}.  Each of them is indeed 
embedded in, and close to the apex of, the associated cloud, as expected in the RDI model.   
We are apparently witnessing an ongoing star formation sequence, with coexistence of  
CTTS, younger PMS associated with BRCs, to the youngest protostars embedded 
within the clouds.

\subsection{RELEVANT TIMESCALES}

In the triggered star-formation scenario, the ages of the concurrent OB stars 
must be longer than the ages of the 
second-generation stars plus the I-front traveling time.  The stars 
$^{1}\theta$~Ori~A, B, C, and D, with spectral types B0.5V, B3V, O6:, and B0.5V, 
respectively \citep{lev76}, are the most massive members in the Trapezium and 
the main sources to create the I-fronts.   The star $^{1}\theta$~Ori~C 
may be somewhat peculiar in its evolutionary status \citep{wal84}, so we adopt 
the age of $^{1}\theta$~Ori~A and D, both of spectral type B0.5V with a lifetime 
of 6$\times$10$^{6}$~yr \citep{sch97} to represent the ages of the OB stars.  

Assuming IC\,2118, LDN\,1616, and the Trapezium are at the same distance from us, 
IC\,2118 and LDN\,1616 are thus separated from the Trapezium by 
about 55~pc.  With an I-front speed, i.e., the sound speed in an H\,{\small II} 
region, $\sim$ 10~km~s$^{-1}$, it takes some 5.5$\times10^{6}$~yr for the I-front to 
traverse.  The lifetimes of CTTS range from a few 10$^{5}$ to a few 10$^{6}$~yr.  
The CTTS associated with the BRCs are younger than those outside the clouds 
(\S 3.2), but even for the oldest CTTS in the region, the combined timescale 
6$\times10^{6}$~yr is consistent with the age of the Trapezium.  

Likewise, the set of timescales can be estimated for the $\lambda$~Orionis system, 
which consists of an O star and an H\,{\small II} region surrounded by an 
unusual molecular cloud ring.  The radius of the molecular cloud ring  
is approximately 2.6~deg, or about 20~pc at 450~pc distance, so it would take 
about 2~Myr for the I-front, originated from $\lambda$~Ori, to travel through 
the H\,{\small II} region.  The age of $\lambda$~Orionis, an O8III 
star with a luminosity of 340,000~L$_{\sun}$ and an effective temperature of 
34,700~K \citep{sch82} can be inferred from evolutionary tracks \citep{sch97} 
to be about 3$\times10^{6}$~yr.  Here again, the age of the triggerer 
$\lambda$~Orionis (3$\times10^{6}$~yr) is consistently longer than the 
expanding timescale of I-front (2$\times10^{6}$~yr) plus the age of the CTTS.

\subsection{SPATIAL DISTRIBUTION OF CLASSICAL T TAURI STARS}

The spatial distribution of CTTS provides telltale clues to the star 
formation history in a molecular cloud.  According to the RDI model, 
small-scale condensations compressed by the I-fronts would reach critical 
equilibria first at the surface layers of BRCs, where the CTTS should be 
found.  If CTTS exist far ahead of the I-front and are still embedded 
in the molecular cloud, the star formation must have 
proceeded in a spontaneous manner, i.e. the embedded CTTS were formed at the 
same time with their massive counterparts.  On the other hand, if CTTS are 
only seen near the surfaces of --- and none embedded inside --- the BRCs, triggered 
star formation is clearly evinced.  In B\,30, B\,35, 
IC\,2118, LDN\,1616, LDN\,1634 and Orion\,East, CTTS are seen only between the massive stars 
and BRCs, and on the compressed (i.e., dense) sides of BRCs 
(see Figure~\ref{fig:images5}, \ref{fig:images6}, and \ref{fig:images7}).  
No CTTS are found further down the compressed regions into the molecular clouds.   

To investigate if the BRCs are indeed void of embedded PMS stars, we have computed 
the probability of our failure to detect them, if they actually existed, because of 
excessive dust extinction in the clouds.  Table~3 contains the extinction 
of IC\,2118, LDN\,1616, LDN\,1634, B\,30, and B\,35.   
Column 1 lists the 5 BRCs.  Columns 2 to 4 are the $E(B-V)$, A$_{V}$, and 
A$_{J}$ in each cloud.  The values of $E(B-V)$ are derived from \citet{sch98}, 
and A$_{V}$/E$_{B-V}$=3.1 and A$_{J}$ = 0.282A$_{V}$ \citep{cox99} 
are used to obtain A$_{V}$ and A$_{J}$ .  If there were CTTS 
embedded in a particular cloud, we would like to estimate how many of them 
would have hidden from our detection.  We created the $E(B-V)$ map \citep{sch98} for 
each cloud, and assumed the same $J$ band luminosities for the embedded 
CTTS as those for the visible CTTS outside the cloud.  We performed Monte Carlo 
computation to distribute randomly the hypothetical CTTS stars on the $E(B-V)$ 
map to check their detectability, given the 2MASS detection limit of $J=15.0$ 
(SNR=30) based on selection criteria discussed in \S~2.  The probability of 
non-detection of embedded CTTS in each cloud is given as Column 5.  As can be 
seen, the probability of CTTS that would have escaped our detection is universally 
very low, $< 0.06$, implying that the extinction in these clouds is too low 
to make any embedded PMS stars invisible.  In other words, if there were 
any embedded sources, we 
would have detected them.  We conclude therefore that the CTTS near the 
surfaces of the BRCs are not part of a young stellar population originally 
embedded in the molecular clouds and later revealed by UV photons from massive 
stars.  They are relics of triggered star formation.  

Most PMS stars are associated with molecular clouds, but some are distinctly 
away from any clouds.  These isolated PMS stars, with their juvenility, could 
not have traversed from the birthplace to their current positions.  They are 
revealed as a consequence of surplus star-forming material being stripped 
off by radiation and winds from nearby massive stars, leaving behind PMS stars away 
from molecular clouds.

\subsection{HIGH STAR FORMATION EFFICIENCY}

Massive stars play a dual role in star formation.  The I-front compression may 
induce birth of stars which otherwise could not have formed.  Alternatively 
molecular clouds may become highly disturbed, hence dispersed, making subsequent 
star formation impossible.  Even if triggered star formation is initiated, the 
working is confined to the outer layers of a cloud.  It is therefore expected 
that star formation may be efficient in localized, compressed regions, but perhaps 
not so in the entire cloud.  The star formation efficiency (SFE) in a BRC can 
be estimated by the ratio of stellar mass to the total mass of stars plus 
cloud.  \citet{dev02} observed the compressed regions of B\,35 and LDN\,1634 in the
HCO$^{+}$ molecular line and determined the total compressed mass of molecular 
clouds to be 27 M$_{\sun}$ for B\,35 and 28 M$_{\sun}$ for LDN\,1634, respectively.  We 
have identified 3 CTTS in each of B\,35 and LDN\,1634, so assuming 0.5--1 M$_{\sun}$ 
for a typical CTTS, the SFEs of the compressed material in B\,35 and in LDN\,1634 
are about 5--10\%, higher than that of a few percent ($<3$\%) in other nearby 
star-forming regions \citep{whi95}.

\section{CONCLUSIONS}

We have developed an empirical set of criteria to select candidate classical T 
Tauri stars on the basis of the infrared colors in the 2MASS Point Source Catalog.
Among the 32 candidates we have obtained the spectra, 24 are found to be bona fide
PMS objects, 4 are M-type stars and the rest 4 are carbon stars, which shows the
effectiveness of our selection method.  The young stellar population provides a
tool to trace ongoing star formation activities.  We find supportive evidence of
triggered star formation in the following bright-rimmed clouds,
B\,35, B\,30, IC\,2118, LDN\,1616, LDN\,1634, and Orion East.

\begin{enumerate}

\item Our data supports the radiation-driven implosion model 
that links the formation of BRCs and of stars.  The ionization fronts from OB stars
compress a nearby cloud, and if the density of the compressed gas exceeds a certain
critical value, new stars would be formed.  Not all BRCs are associated with CTTS.
Only BRCs associated with strong IRAS 100~$\micron$ emission (tracer of high density)
and H$\alpha$ emission (tracer of ionization front) show signs of ongoing star formation.

\item We find that CTTS closer to BRCs are progressively younger, a configuration 
consistent with the sequential process of triggered star formation.  
The brightest CTTS, spatially closest to BRCs, are the young stellar objects
which just reveal themselves on the birthline and begin to descend down the Hayashi tracks.

\item The ages of the concurrent OB stars are comparable or older than the ages of the
next-generation stars plus the I-front traveling times.  These timescales are consistent
with the causality, or sequential, nature of the star formation process.

\item CTTS are found only in between the OB stars and the molecular clouds with which
the ionization fronts interact.  There are no CTTS ahead of the ionization fronts into the clouds.
The lack of embedded CTTS cannot be due to detection sensitivity because the
extinction of these BRCs is generally low.

\item Current star formation in the BRCs takes place in the compressed outer 
layers of the molecular clouds.  The relatively enhanced star formation efficiency 
suggests a triggering---as opposed to a spontaneous---star formation process.

\end{enumerate}

\acknowledgments

This research makes use of data products from the Two-Micron All-Sky Survey, 
which is a joint project of the University of Massachusetts and the Infrared 
Processing and Analysis Center/California Institute of Technology, funded by the 
National Aeronautics and Space Administration and the National Science Foundation.  
We also use the Southern H-Alpha Sky Survey Atlas (SHASSA), supported by the 
National Science Foundation.  
We are grateful to the anonymous referee for helpful suggestions that improve the 
quality of his paper.  We also thank the staff at the Beijing Astronomical Observatory
for their assistance during our observing runs.  Hsu-Tai Lee wants to thank Paul Ho 
for valuable discussions.  We acknowledge the financial support of the grants 
NSC92-2112-M-008-047 of the National Science Council, and 92-N-FA01-1-4-5 of the 
Ministry of Education of Taiwan.

\clearpage

\clearpage

\clearpage

\clearpage

\clearpage

\clearpage

\clearpage

\clearpage

\clearpage

\clearpage

\begin{deluxetable}{ll}

\tablecaption{Selection Criteria of CTTS from 2MASS Point Source Catalog}
\tabletypesize{\footnotesize}
\rotate
\tablenum{1}

\tablehead{\colhead{Criterion} & \colhead{Description} \\}
\startdata
Photometric quality flag $(ph\_qual) = AAA$ & 
$SNR \geq 10$ and $[jhk]\_cmsig < 0.10857$ for  \\
  & best-quality detections in photometry and astrometry \\
Blend flag $(bl\_flg) = 111$ & No source blending  \\
Contamination and confusion flag $(cc\_flg) = 000$ & Source unaffected by
known artifacts \\
Proximity (prox) $>$ 5.0\arcsec & Photometry not confused by nearby objects \\
Minor Planet Flag $(mp\_flg) = 0$ & Not a minor planet \\
$(j\_m-j\_m\_stdap)+(h\_m-h\_m\_stdap)$ & 
To distinguish an extended source from a point source  \\
$+(k\_m-k\_m\_stdap) \leq 0.3 mag$  &  \\
$(j\_m-h\_m)-1.7(h\_m-k\_m)+0.0976 \leq 0$ & 2MASS colors of CTTS \\
$(j\_m-h\_m)-1.7(h\_m-k\_m)+0.4500 \geq 0$ \\
$(j\_m-h\_m)-0.493(h\_m-k\_m)-0.439 \geq 0$ \\
Optical counterpart (a) $\neq$ 0 & With an optical counterpart \\
signal-to-noise ratio ($[jhk]\_snr$) $> 30$ & SNR $>$ 30 \\
\enddata

\end{deluxetable}

\clearpage

\begin{deluxetable}{lcrrrcccc}

\tablecaption{Spectroscopic Observations of Young Star Candidates}
\tabletypesize{\footnotesize}
\rotate
\tablenum{2}

\tablehead{\colhead{Star} & \colhead{2MASS} & \colhead{$J$} & \colhead{$H$} & \colhead{$K_{S}
$} & \colhead{Sp.} & \colhead{Region} &
\colhead{Forbidden Lines} & \colhead{Other Name} \\
\colhead{} & \colhead{} & \colhead{} & \colhead{} &
\colhead{} & \colhead{Type$^{a}$} & \colhead{Region} & \colhead{Line$^{b}$} & \colhead{} \\
\colhead{} & \colhead{} & \colhead{(mag)} & \colhead{(mag)} &
\colhead{(mag)} & \colhead{} & \colhead{} & \colhead{} & \colhead{} }

\startdata
1 & J05350900-0427510 & 10.797 & 10.109 & 9.634 & K0 & Orion A & - & [R2001] 1754 \\
2 & J05353672-0510004 & 13.309 & 12.481 & 11.936 & M3 & Orion A & -  & [AD95] 1868 \\
3 & J05385149-0801275 & 12.967 & 11.525 & 10.513 & M5 & Orion A & O, S \\
4 & J05402496-0755353 & 12.446 & 11.367 & 10.658 & M4 & Orion A & -  \\
5 & J05412534-0805547 & 10.604 & 9.824 & 9.287 & Cont. & Orion A & O & HBC 181 \\
6 & J05412771-0931414 & 12.147 & 11.033 & 10.272 & Cont. & Orion A & O & Kiso A-1048 45 \\
7 & J05423584-0958552 & 9.973 & 8.959 & 8.227 & Cont. & Orion A & -  \\
8 & J05432701-0959375 & 10.984 & 10.166 & 9.542 & K5 & Orion A & -  \\
9 & J05342764-0457051 & 13.737 & 12.890 & 12.286 & M2 & Orion A & -  & [CHS2001] 3990 \\
10 & J05342978-0451477 & 13.241 & 11.781 & 10.910 & K7 & Orion A & -  & [CHS2001] 4167 \\
11 & J05122053-0255523 & 10.425 & 9.688 & 9.140 & K3 & - & -  & V531 Ori \\
12 & J05371885-0020416 & 11.577 & 10.761 & 10.263 & K5 & - & -  & Kiso A-0904 58 \\
13 & J06014515-1413337 & 11.698 & 10.896 & 10.211 & K5 & - & -  & IRAS 05594-1413 \\
14 & J05073016-0610158 & 10.822 & 9.572 & 8.611 & M0 & IC 2118 & O, S \\
15 & J05073060-0610597 & 10.120 & 9.040 & 8.251 & M2 & IC 2118 & O \\
16 & J05202573-0547063 & 11.036 & 10.202 & 9.608 & Cont. & LDN 1634 & O, S & V534 Ori \\
17 & J06075243-0516036 & 10.417 & 9.474 & 8.639 & K0 & Mon R2 &  & HBC 518 \\
18 & J06075463-0614342 & 11.676 & 10.851 & 10.094 & K0 & Mon R2 & - \\
19 & J06080003-0519022 & 11.384 & 10.413 & 9.649 & K0 & Mon R2 & - \\
20 & J06261259-1028346 & 12.412 & 11.352 & 10.487 & Cont. & LDN 1652 & O, S \\
21 & J06265529-0958014 & 11.914 & 10.901 & 10.126 & Cont. & LDN 1652 & O \\
22 & J06265731-0959395 & 13.161 & 12.137 & 11.363 & Cont. & LDN 1652 & O \\
23 & J06273428-1002397 & 12.522 & 11.737 & 11.176 & K7 & LDN 1652 & -  \\
24 & J06280028-1003420 & 12.006 & 10.935 & 10.091 & Cont. & LDN 1652 & O \\
25 & J05375005-1548114 & 6.573 & 5.571 & 4.912 & M9 & - & -  & IRAS 05355-1549 \\
26 & J06222376-0255509 & 10.410 & 8.734 & 7.559 & C & - & -  & IRAS 06198-0254 \\
27 & J06380179-0557170 & 8.000 & 6.711 & 5.841 & C & - & -  \\
28 & J06270654-0540512 & 8.082 & 6.751 & 5.882 & C & - & -  \\
29 & J06160006-1727270 & 10.213 & 8.978 & 8.040 & C & - & -  \\
30 & J06265037-0738456 & 9.830 & 8.845 & 8.543 & M2 & - & -  \\
31 & J05345877-0928274 & 7.700 & 6.818 & 6.295 & M9 & - & -  & V653 Ori \\
32 & J05413322-0755022 & 12.209 & 11.174 & 10.526 & M2 & - & -  & Haro 4-488 \\
\enddata

\tablenotetext{a}{Cont.--continuum spectra; C--carbon star}
\tablenotetext{b}{O--[O\,{\scriptsize I}] ;S--[S\,{\scriptsize II}]}

\end{deluxetable}

\clearpage

\begin{deluxetable}{lcccc}

\tablecaption{EXTINCTION OF BRIGHT-RIMMED CLOUDS}

\tablenum{3}

\tablehead{\colhead{BRC} & \colhead{$E(B-V)^{a}$} & \colhead{A$_{V}(R=3.1)$} & \colhead{A$_{J}$} & \colhead{Nondetection Prob.} \\ 
\colhead{} & \colhead{(mag)} & \colhead{(mag)} & \colhead{(mag)} & \colhead{} } 

\startdata

IC 2118  & 1.1--0.2 & 3.4--0.6 & 0.96--0.17 & 0.021 \\
LDN 1616 & 2.3--0.2 & 7.1--0.6 & 2.01--0.17 & 0.051 \\
LDN 1634 & 1.1--0.3 & 3.1--0.9 & 0.87--0.25 & 0.013 \\
B 30     & 2.6--1.5 & 8.1--4.7 & 2.28--1.31 & 0.058 \\
B 35     & 1.3--0.5 & 4.0--1.6 & 1.14--0.44 & 0.001 \\

\enddata

\tablenotetext{a}{\citet{sch98}}

\end{deluxetable}

\end{document}